\documentclass[usenatbib,useAMS]{mnras}
\usepackage{times}
\usepackage{color}
\usepackage{color}
\usepackage{mathrsfs}
\usepackage{multicol}
\usepackage{bm}		
\usepackage{pdflscape}	
\usepackage{amsfonts,amsmath,amssymb,mathrsfs}
\usepackage{graphicx,epsf,epsfig}
\usepackage{inputenc}
\usepackage{bm}
\usepackage{longtable}
\usepackage[usenames,dvipsnames]{xcolor}
\usepackage{multirow}
\usepackage{hyperref}
\usepackage{times}
\usepackage{color}
\usepackage{tabularx}
\usepackage{appendix}

\def\be{\begin{equation}}
\def\ee{\end{equation}}
\def\bea{\begin{eqnarray}}
\def\eea{\end{eqnarray}}

\def\prd{Phys. Rev. D}

\def\mnras{Mon. Not. Roy. Astr. Soc.}
\def\apj{Astrophys. J.}
\def\apjl{Astrophys. J. Lett.}

\def\aap{Astron. Astrophys.}

\def\aapr{Astron. Astrophys. Rev.}

\def\araa{Annual Rev. of Astron. Astrophys.}

\def\pasj{Publ. Astr. Soc. Japan }

\def\physrep{Phys. Rep.}
\def\jcap{J.C.A.P.}

\title{Numerical analyses of M31 dark matter profiles }

\author[K.~Boshkayev, et. al.]{Kuantay~Boshkayev,$^{1,2,3}$\thanks{kuantay@mail.ru} Talgar~Konysbayev,$^{1,2}$\thanks{talgar\_777@mail.ru} Yergali~Kurmanov,$^{1,2}$\thanks{kurmanov.yergali@kaznu.kz} Orlando~Luongo,$^{2,4,5}$\thanks{orlando.luongo@unicam.it}\newauthor  Marco~Muccino,$^{2}$\thanks{marco.muccino@lnf.infn.it}
Hernando~ Quevedo,$^{2,6,7}$\thanks{quevedo@nucleares.unam.mx} and Gulnur~ Zhumakhanova$^{8}$\thanks{zhumakhanovag@gmail.com}\\
$^1$National Nanotechnology Laboratory of Open Type,  Al-Farabi ave. 71, Almaty, Kazakhstan.\\
$^2$Al-Farabi Kazakh National University, Al-Farabi ave. 71, Almaty, Kazakhstan.\\
$^3$Department of Physics, Nazarbayev University, 53 Kabanbay Batyr, Astana, Kazakhstan.\\
$^4$Scuola di Scienze e Tecnologie, Divisione di Fisica, Universit\`a di Camerino, Via Madonna delle Carceri 9, Camerino, Italy.\\
$^5$Istituto Nazionale di Fisica Nucleare, Sezione di Perugia, via A. Pascoli, I-06123 Perugia, Italy. \\
$^6$Instituto de Ciencias Nucleares, Universidad Nacional Aut\'onoma de M\'exico, Mexico. \\
$^7$Dipartimento di Fisica and ICRA, Universit\`a di Roma “La Sapienza”, Roma, Italy.\\
$^8$National Center of Science and Technology Evaluation, 221 Bogenbay batyr, Almaty, Kazakhstan.
}
\begin{document}

\maketitle

\begin{abstract}
We reproduce the rotation curve of the  Andromeda galaxy (M31) by taking into account its bulge, disk, and halo components, considering the last one to contain the major part of dark matter mass. Hence, our prescription is to  split the galactic bulge into two components, namely, the inner and main bulges, respectively. Both bulges are thus modeled by exponential density profiles since we underline that the widely accepted de Vaucouleurs law fails to reproduce the whole galactic bulge rotation curve. In addition, we adopt various well-known phenomenological dark matter profiles to estimate the dark matter mass in the halo region. Moreover, we apply the least-squares fitting method to determine from the rotation curve the model free parameters, namely, the characteristic (central) density, scale radius, and consequently the total mass. To do so, we perform Markov chain Monte Carlo statistical analyses based on the Metropolis algorithm, maximizing our likelihoods adopting velocity and radii data points of the rotation curves. We do not fit separately the components for bulges, disk and halo, but we perform an overall fit including all the components and employing all the data points. Thus, we critically analyze our corresponding findings and, in particular, we employ the Bayesian Information Criterion  to assess the most accredited model to describe M31 dark matter dynamics.
\end{abstract}


\section{Introduction}\label{sec:intro}

At large distances from the centers, stars within spiral galaxies rotate with anomalous speeds \citep{Banik,Elson}, indicating that galaxy rotation curves are fundamental to disclose the unknown nature of \emph{dark matter} (DM).
The evidences of large amounts of invisible matter, distributed differently from the stellar and gaseous disks, turned up in the 1970s \citep{Roberts1978,Faber1979,Rubin1980,Bosma1981}. In fact, optical and 21-cm rotation curves (RCs) were found to behave in an anomalous way, fully-incompatible with the Keplerian fall-off predicted from the outer distribution of luminous matter. Hence, the determination of the DM abundance, \textit{i.e.}, of the mass distribution in spiral galaxies, is essential in order to figure out its physical properties. In this regard, a widely-consolidate technique consists in  analyzing rotation curves \citep{Sofue2009,Salucci:2018hqu,Dehghani:2020cvl}.

Furthermore, the corresponding DM gravity field keeps galaxies and cluster of galaxies stable, being essential to describe perturbations regimes in cosmology too \citep{Bernardeau:2001qr}. Confirmations of the  DM existence are numerous and come from several observational evidences, \textit{e.g.}, in galaxy clusters, hot gas motion, gravitational microlensing \citep{Nakama}, and clustering of structures \citep{EHTC}.

However, the DM nature is still highly-debated since no suitable candidates as  DM particles have been experimentally found. In particular, the existence of DM appears evident through dynamical  measurements,  but not with  interaction with ordinary matter or radiation, providing that no ground-based laboratories have been able to detect DM particles  yet \citep{Woithe}. Consequently, the DM nature is still challenging and unknown \citep{Herrera}. In this respect, a widely-accepted hypothesis is that DM consists of a class of weakly interacting massive particles and/or of (ultra-)light particles, although alternative viewpoints have been  proposed\footnote{Geometric DM has been formulated in the context of extended theories of gravity, e.g., \citet{Capozziello:2019cav}, DM quasi-particles in \citet{Belfiglio:2022cnd} and/or unified dark sector models \citep{Boshkayev:2019qcx}, modifications of the Newton law in \citet{Milgrom:2019cle}, and so on.}, see \textit{e.g.} \citep{Bertone2018} and references therein\footnote{Possible scenarios that minimally extend the standard model of particle physics would predict axions, gravitinos, etc., see e.g. \citet{Bertone2005,Salucci2019,Quiskamp2022,2022PTEP.2022a3F01M}. Other alternatives involve highly-massive structures, such as MACHOS, see e.g. \citet{Brandt2016,Chapline2016,Bai2020,Katz2020,Yang2022PhRvD}, or modifications of gravity at large scales due to extra terms, see e.g. \citet{Capozziello:2019cav}. }.

In addition, at zeroth approximation \emph{different} phenomenological density profiles are adopted to explain the observed rotation curves in galaxies. For instance, for the Milky Way (MW) galaxy, different DM profiles are used to obtain information about the RCs from the central part of the galaxy to the halo~\citep{2021MNRAS.508.1543B}.

In this work, motivated by the fact that detailed RC outcomes have been investigated in the Andromeda galaxy~\citep{Corbelli, Carignan, 1970ApJ...159..379R}, and are used for the distribution of masses in the disk and the dark halo, we analyze the M31 RCs and consider all the possible morphological configurations of the galaxy itself. In particular, we focus on both inner and main bulge regions, on the disk and finally on the halo component, considered as the main target whose phenomenological profile, among the consolidate profiles, is preferable from a statistical standpoint. To this end, we work out six profiles, \textit{i.e.}, the \emph{exponential sphere (hereafter ES)},  \emph{pseudo-isothermal} (hereafter ISO),  \emph{Burkert}, \emph{Navarro-Frenk-White} (NFW),  \emph{Moore}, \emph{Beta} and the \emph{Brownstein profile} and we estimate the corresponding DM abundance in each component of the galaxy. Afterwards, we follow the subsequent steps:
\begin{itemize}
    \item[-] analysis of the RC data of M31,
    \item[-] determination of the model free parameters from the log-likelihood approach and, finally,
    \item[-] reconstruction of the theoretical RC of M31.
\end{itemize}

The latter step is attained by using the data from \citet{Sofue2015} and carrying out a similar analysis that differs by the following aspects:
\begin{enumerate}
    \item[-] we divide the bulge into \textit{inner} and \textit{main bulges}.
A similar analysis has been performed for the MW \citep[see e.g.,][]{Sofue2013}, where the RC is better described in terms of two bulges, and by analogy can be applied to M31 since two distinct bumps below 4 kpc are clearly visible in the RC\footnote{
The bulge structure greatly varies with the type of galaxies. For example, elliptical galaxies consist only of a spherical bulge, whereas irregular ones have a weak bulge.
Most of the RCS of the galaxies considered by  \citet{Bernal2018} exhibits only one bulge. However, some galaxies may have a less evident second bulge. Indeed, according to \citet{2007ApJ...658L..91B}, \citet{2013PASA...30...27M} and \citet{2017MNRAS.466.4279B} M31 has two bulges.};
    \item[-] we adopt the exponential density profile for inner and main bulges in analogy to \citet{Sofue2013} and \citet{2021MNRAS.508.1543B}.
    \item[-] we consider alternative profiles for the halo and determine the most suitable from the statistical point of view.
\end{enumerate}

In this respect, we extend the analysis performed in \citet{Sofue2015}, where the RC has been constructed for a bulge, a disk and a dark halo, employing only the NFW profile for the halo. Remarkably, we demonstrate that the NFW is not the most accredited profile under the above hypotheses. Further, we compare our findings with the MW, emphasizing the main physical differences.

The paper is organized as follows. In Section~\ref{sec:profiles}, we review well-known phenomenological DM profiles. In Section~\ref{sec:mass}, we introduce the basic ideas and hypotheses of the work and present the results of fitting RC data for M31 and MW galaxies.  In Section~\ref{sec:concl}, we briefly discuss our outcomes and compare them  with other known results. Finally, we summarize the main conclusions of our work.


\section{Phenomenological dark matter profiles}\label{sec:profiles}

Bulge and disk mainly consist of baryonic (visible) matter, with the bulge mostly composed of older stars and the disk composed of star forming regions, gas and dust.

Galactic halos are assumed to be composed of DM only. Their density profiles can be obtained by using numerical simulation methods of stars dynamics in galaxies.
Here, we selected the most adopted ones, namely ISO, Beta, Burkert, Brownstein, Moore, NFW and ES. All these profiles are characterized by two parameters: the characteristic central density of DM $\rho_{0}$ and the scale radius $r_0$, that can be expressed as functions of the dimensionless distance $x=x(r)=r/r_0$.
The above models, split into:\\

\emph{{\bf \emph{Cored} profiles.}}
\begin{enumerate}
\item[{\bf 1.}] \emph{ISO}~\citep{Jimenez}
\begin{equation}
\label{eq:sample1}
\rho_{ISO}(x) = \frac{\rho_0}{1+x^2}\,;
\end{equation}
\item[{\bf 2.}] \emph{Burkert}~\citep{Burkert}
\begin{equation}
\label{eq:sample2}
\rho_{Bur}(x) = \frac{\rho_0}{\left(1+x\right)\left(1+x^2\right)};\
\end{equation}
\item[{\bf 3.}] \emph{Beta with $\beta=1$}~\citep{1995MNRAS.275..720N,2020Galax...8...37S}
\begin{equation}
\rho_{Beta}(x) = \frac{\rho_0}{\left(1+x^2\right)^{1.5}}\,;
\label{eq:sample3}
\end{equation}
\item[{\bf 4.}] \emph{Brownstein} ~\citep{2006ApJ...636..721B,2020Galax...8...37S}
\begin{equation}
\rho_{Bro}(x) = \frac{\rho_0}{1+x^3}.\
\label{eq:sample4}
\end{equation}
\end{enumerate}
\,\\
\emph{{\bf \emph{Cuspy} profiles.}}
\begin{enumerate}
\item[{\bf 5.}] \emph{NFW} \citep{Navarro}, based on cosmological models of halo formation,
\begin{equation}
\rho_{NFW}(x) = \frac{\rho_0}{ x\left(1+x\right)^2}\,;
\label{eq:sample5}
\end{equation}
\item[{\bf 6.}] \emph{Moore} \citep{Moore}
\begin{equation}
\rho_{Moore}(x) = \frac{\rho_0}{ x^{1.16}\left(1+x\right)^{1.84}}\,.
\label{eq:sample6}
\end{equation}
\end{enumerate}

The distinction between cored and cuspy models concerns the  cusp of density as $r\rightarrow0$ for the latter two approaches and the smoothness of the first four models in the same regime.
\begin{figure}
\centering
\includegraphics[width=0.97\linewidth]{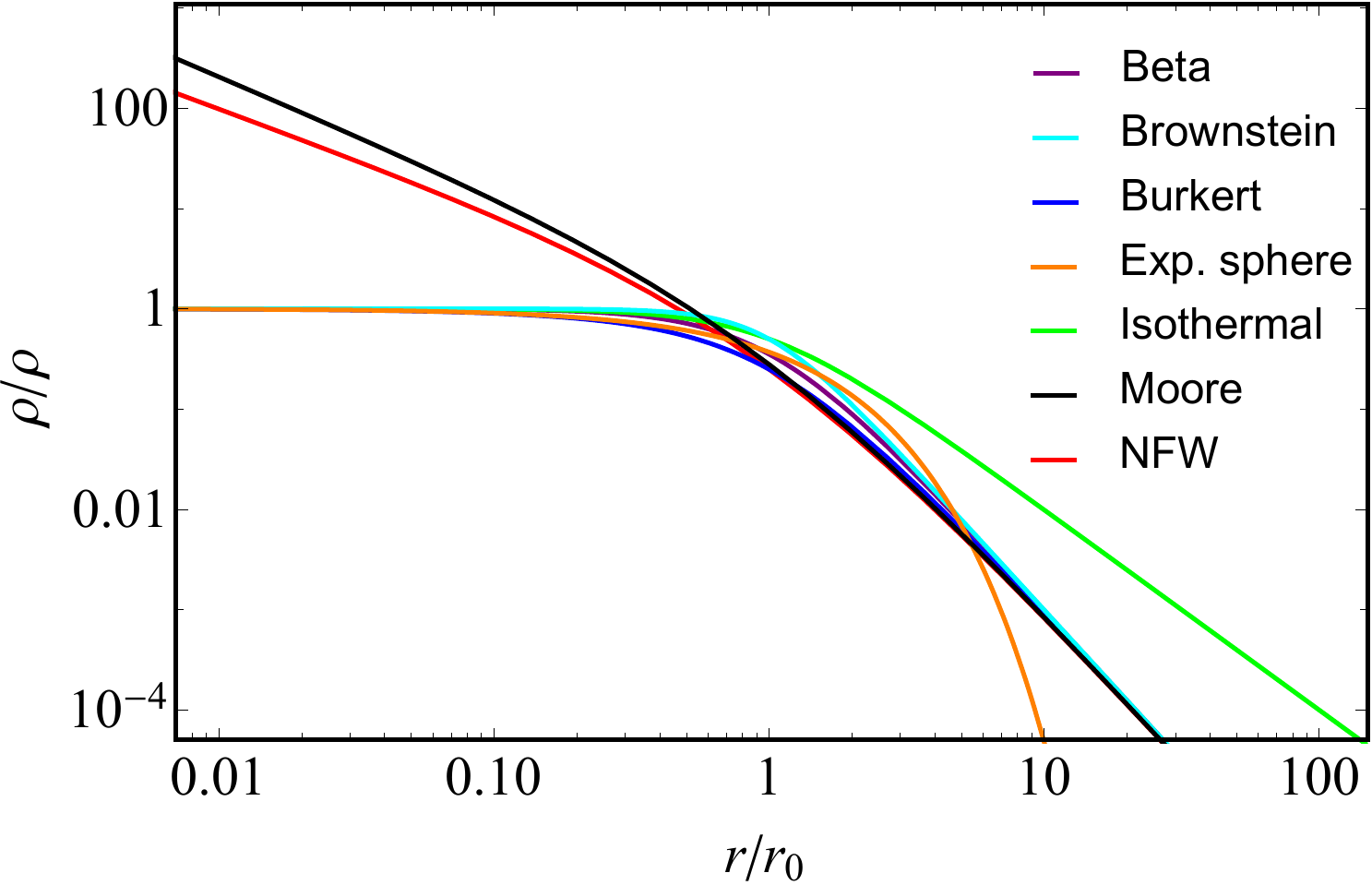}
\caption{Comparison among phenomenological DM density profiles. Here, $\rho_0$ is not fixed since we plot the normalised density $\rho/\rho_0$.}
\label{fig:profile}
\end{figure}

Figure~\ref{fig:profile} shows the dependence of $\rho/\rho_0$ on $r/r_0$ for the DM phenomenological profiles. In this dimensionless representation, one can see that the profiles differ from each other at small and large distances, as for each profile one can have different values of $\rho_0$ and $r_0$. Any profile generates a distinct RC and hence yields corresponding DM mass distributions that can be calculated as
\begin{equation}\label{eq:massprof}
M_h(r)=\int_0^r4\pi x^{2}\rho(x)dx\,,
\end{equation}
where $\rho(x)$ is any of the DM profiles from Eqs.~\eqref{eq:sample1}--\eqref{eq:sample6}.

\subsection{Inner parts of M31}

Inner parts of galaxies are usually modeled by the de Vaucouleurs profile defining the surface mass density
\be
\label{eq-smdb}
\Sigma_{\rm b}(x) = \Sigma_{bc} \exp \left[\kappa \left(1-x^{1/4}\right)\right],
\ee
where $\kappa=7.6695$, $x=r/r_b$, $\Sigma_{bc}$ is the central value of the surface mass density, and $r_b$ is the scale radius of the bulge.
The total mass of the bulge is given by
\be
M_{bt}= 2 \pi \int_0^\infty \Sigma_{\rm b}(r)\,r\,dr =\eta r_b^2 \Sigma_{ bc},
\ee
where $\eta=22.665$ is a dimensionless constant.
The mass enclosed within a sphere of radius $r$ is given by
\be
M_{bt}(r)= 4 \pi \int_0^r \rho_{\rm b}(y)\,y^2dy,
\ee
where the volume density $\rho_{\rm b}(y)$ is calculated as
\be
\rho_{\rm b}(y) = \frac{1}{\pi} \int_y^\infty \frac{d\Sigma_{\rm b}(x)}{dx}\frac{dx}{\sqrt{x^2-y^2}}.
\ee

It was shown that the de Vaucouleurs law fails to fit the bulge RC of the MW galaxy. As an alternative in \citet{Sofue2013} the bulge was split into inner and main bulges, both modeled by ES profiles
\be
\label{eq:sample15}
\rho(x)=\rho_0 \exp (-x)\,,
\ee
where $x=r/r_0$. The mass enclosed within a sphere of radius $r$ reads
\be
M(r) = M_0 \left[1-\exp(-x)\left(1+x+x^2/2\right)\right]\,,
\ee
where the total mass $M_0$ is given by
\be
M_0=4 \pi\int_0^\infty  r^2 \rho(r) dr=8 \pi r_0^3 \rho_{0}\,.
\ee

Later in the text, we show that inner and main bulges modeled by two profiles as in Eq. \eqref{eq:sample15} are statistically and visually better than the case modeled by one bulge only, adopting Eq. \eqref{eq-smdb}. In addition to this prescription, we underline that Eq. \eqref{eq-smdb} represents a surface density, whereas Eq. \eqref{eq:sample15} consists of a volume density and, consequently, we cannot compare the two profiles to each other.

\subsection{The galactic disk}

The galactic disk is usually represented by an exponential disk with a surface mass density \citep{1970ApJ...160..811F}
\be\label{eq-smdd}
\Sigma_{\rm d} (x)=\Sigma_{dc} \exp(-x)\,,
\ee
where $x=r/r_d$, $\Sigma_{dc}$ is the central value of the surface density and $r_d$ is the disk scale radius.
The mass enclosed within a radius $r$ is
\be
M_{d}(r)= M_{dt} \left[1-\exp(-x)\left(1+x\right)\right]\,,
\ee
where the total mass is $M_{dt}$ is given by
\be
M_{dt} = 2 \pi \int_0^\infty \Sigma_{\rm d}(r)\,r\,dr = 2\pi r_d^2 \Sigma_{dc}\,.
\ee
The rotation curve for a thin exponential disk is given by \citep{1987gady.book.....B}.
\be
\label{expdisk}
V_{\rm d}(y)= \sqrt{\frac{2G M_{dt}}{r_d} y^2 \left[I_0(y)K_0(y)-I_1(y)K_1(y)\right]}\,,
\ee
where $y=r/(2r_{\rm d})$, and $I_i$ and $K_i$ are the modified Bessel functions of the first and second kind, respectively.


\section{Inference of the model parameters and dark matter mass }\label{sec:mass}

The linear velocity $V$ of stars and gas depends upon the galacto-centric distance $r$ and it is determined by equating the centripetal and gravitational forces (via the internal gravitational potential $\Phi$) acting on a star moving in a circular orbit
\begin{equation}\label{eq:sampleF}
V(r)=\sqrt{\frac{GM(r)}{r}}=\sqrt{r\frac{\partial \Phi(r)}{\partial r}},\,
\end{equation}
where $G$ is the gravitational constant.

Now, we can fit the RC of M31 to extract the model free parameters.
We here consider the case with one bulge, disk and halo, in analogy with \citet{Sofue2015}, and the case with the splitting of the bulge into inner and main components, in agreement with \citet{Sofue2013}.
In the latter case, the resulting RC is given by
\begin{equation}\label{eq:theor_rc}
V(r)^{2}=V_{ib}(r)^{2}+V_{mb}(r)^{2}+V_{d}(r)^{2}+V_{h}(r)^{2},
\end{equation}
where $V_{ib}, V_{mb}, V_{d}$ and $V_{h}$ are the linear velocities of test particles (stars) in the gravitational field of the mass distribution at distance $r$ of inner and main bulges, disk, and halo, respectively.

We modified the \texttt{Wolfram Mathematica} code from \citet{2019PhRvD..99d3516A} and performed Markov chain -- Monte Carlo fits of the RC of Andromeda, by means of the Metropolis algorithm, searching for the best-fit parameters maximizing the log-likelihood
\begin{equation}
\label{loglike}
    \ln \mathcal{L} = -\frac{1}{2}\sum_{k=1}^{N}\left\{\left[\dfrac{V_k-V(r_k)}{\sigma V_k}\right]^2 + \ln(2\pi\,\sigma V_k^2)\right\}\,,
\end{equation}
where $N=46$ are the velocity data points $V_k$ at radii $r_k$ of the RC and $\sigma V_k$ are the attached errors.
We first reproduced the results of \citet{Sofue2015} and then inferred the model free parameters using Eq.~\eqref{eq:theor_rc}.
Differently from \citet{Sofue2013,Sofue2015}, we do not split the RC and fit separately the components for bulge(s), disk and halo, but we perform an overall fit including all the components and employing all the data points without splitting them.

To assess the best-fit model out of the six profiles when considering the halo, we used the Bayesian Information Criterion (BIC)
\begin{equation}
\label{eq:sampleBIC}
{\rm BIC}=-2\ln \mathcal{L}_{\rm max}+k\ln N\,,
\end{equation}
where $k$ is the number of model parameters and $\mathcal{L}_{\rm max}$ the maximum value of the log-likelihood \citep{Yunis}.
In general, the model with the lowest BIC test, say ${\rm BIC}_0$, is considered to be the reference (best-suited) model. The statistical evidence in support of the reference model to  the best-fitting one, when compared to other models, is certified by the difference $\Delta{\rm BIC}={\rm BIC}-{\rm BIC}_0$ and, in particular, by:
\begin{itemize}
    \item[1.] $\Delta{\rm BIC}\in[0,\,2]$ showing a weak evidence.
    \item[2.] $\Delta{\rm BIC}\in (2,\,6]$ showing a mild evidence.
    \item[3.] $\Delta{\rm BIC}>6$ showing a huge evidence.
\end{itemize}

\section{Theoretical results}

\begin{figure}
\centering
\includegraphics[width=0.97\linewidth]{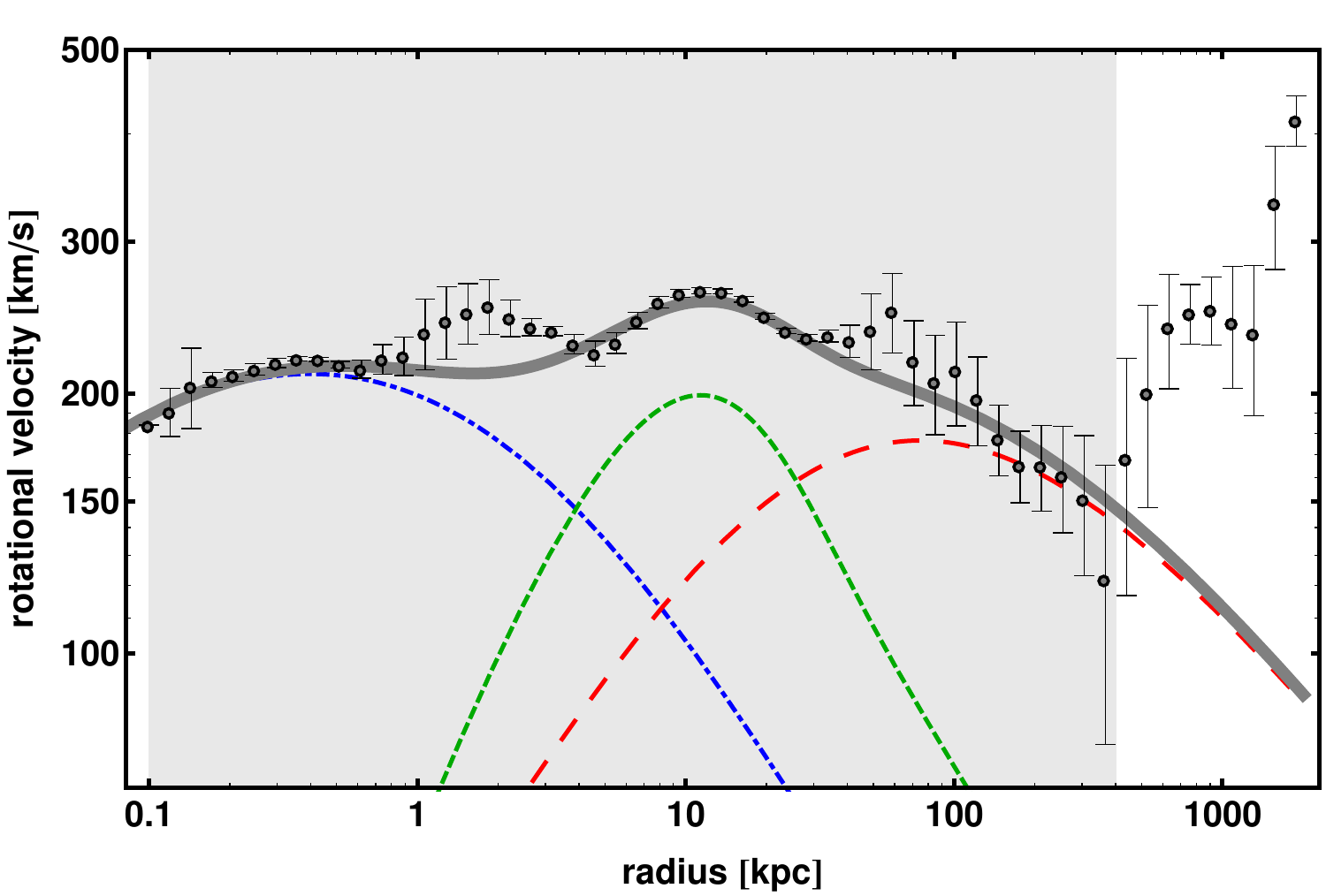}
\caption{Best-fit RC of M31 (thick, gray) composed of bulge (de Vaucouleurs, dot-dashed blue), disk (Freeman, dashed green) and halo (NFW, long-dashed red) components. The shaded area marks the data (black dots with error bars) considered for the fitting. Reproduced from \citet{Sofue2015}.}
\label{fig:M31_Sofue}
\end{figure}

Fig.~\ref{fig:M31_Sofue} and Tables~\ref{tab:first} reproduce the results of the RC analysis of M31 obtained by \citet{Sofue2015}.
The bulge is modelled by the de Vaucouleurs profile \citep{1958ApJ...128..465D}, the disk is approximated by the exponential disk model \citep{1970ApJ...160..811F} and the halo was described by the NFW model \citep{Navarro}. The fitting was performed within the range $0$--$20$ kpc for the bulge, $0$--$40$ kpc for the disk, and $0$--$385$~kpc for the halo \citep{Sofue2015}.
\begin{table}
\centering
\caption{The best-fit parameters of M31 from \citet{Sofue2015}. The halo model  is the NFW profile and its mass was calculated within $r=200$ kpc.}
\label{tab:first}
\begin{tabular}{cccc}
\hline
Parts        & Total mass              & Scale radius      & Central density\\
             &  ($10^{11}M_{\odot}$)   &  (kpc)            &  ($10^{-3} M_{\odot}$/pc$^3$) \\
\hline
Bulge        & $0.35\pm 0.004$         & $1.35\pm 0.02$    & --- \\

Disk         & $1.26\pm 0.08$          & $5.28\pm 0.25$    & ---  \\

Halo         & $12.30\pm 2.60$.        & $34.60\pm 2.10$   & $2.23\pm 0.24$  \\
 \hline
\end{tabular}
\end{table}

Table~\ref{tab:second} lists the best-fit results obtained from the $\sim0$--$400$~kpc RC by considering a single bulge \citet{Sofue2015}. We fixed the de Vaucouleurs profile for the bulge, the Freeman profile for the disk, whereas for the halo we explored the profiles in Eqs.~\eqref{eq:sample1}--\eqref{eq:sample6} and the ES profile.
This comparison among all the DM halo profiles allows us to perform better and objective analyses.
The last combination of models listed in Table~\ref{tab:second} represents a direct comparison, though performed with a different methodology (see Sec.~\ref{sec:mass}), with the results of \citet{Sofue2015} summarized Table~\ref{tab:first}. As one can see, the NFW profile is not the most preferred model according to $\Delta$BIC.

\begin{table*}
\setlength{\tabcolsep}{1.6em}
\renewcommand{\arraystretch}{1.4}
\centering
\caption{Best-fit RC parameters of M31 obtained by considering one bulge (de Vaucouleurs), disk (Freeman) and the halo models considered in this work. The halo mass has been calculated within $r=200$~kpc. The $\Delta{\rm BIC}$ value is computed with respect to the minimum BIC in Table~\ref{tab:third}.}
\begin{tabular}{cccccccr}
\hline\hline
\multicolumn{2}{c}{Bulge}		& \multicolumn{2}{c}{Disk}		& \multicolumn{3}{c}{Halo}  &  \multirow{3}{*}{$\Delta {\rm BIC}$}\\
\cline{1-7}
$M_{b}$	&	$r_{b}$	&	$M_{d}$	&	$r_{d}$&	$\rho_0$		&	$r_0$	&	$M_h$	&\\
($10^{11}$~M$_\odot$) & (kpc) & ($10^{11}$~M$_\odot$) & (kpc) & ($10^{-3}$~M$_\odot$/pc$^3$) & (kpc) & ($10^{11}$~M$_\odot$) &\\
\hline
\multicolumn{2}{c}{de Vaucouleurs}		& \multicolumn{2}{c}{Freeman}		& \multicolumn{3}{c}{Beta} &  \\
$0.43_{-0.02}^{+0.05}$ & $1.58_{-0.08}^{+0.16}$ & $1.54_{-0.07}^{+0.10}$ & $4.98_{-0.20}^{+0.27}$ & $3.10_{-0.70}^{+0.48}$ & $26.23_{-2.08}^{+3.32}$ & $12.22^{+4.23}_{-3.64}$  &  $21.48$\\
\hline
\multicolumn{2}{c}{de Vaucouleurs}      & \multicolumn{2}{c}{Freeman}		& \multicolumn{3}{c}{Brownstein}  &  \\
$0.45_{-0.03}^{+0.04}$ & $1.64_{-0.10}^{+0.13}$ & $1.66_{-0.04}^{+0.07}$ & $5.21_{-0.15}^{+0.24}$ & $1.82_{-0.28}^{+0.13}$ & $30.12_{-1.03}^{+3.56}$ & $11.82^{+3.56}_{-2.07}$ & $19.43$ \\
\hline
\multicolumn{2}{c}{de Vaucouleurs}      & \multicolumn{2}{c}{Freeman}		& \multicolumn{3}{c}{Burkert}  &  \\
$0.43_{-0.03}^{+0.05}$ & $1.56_{-0.09}^{+0.17}$ & $1.45_{-0.09}^{+0.09}$ & $4.87_{-0.20}^{+0.21}$ & $5.37_{-1.32}^{+1.14}$ & $22.93_{-2.13}^{+3.62}$ & $12.16^{+5.30}_{-4.04}$ & $22.68$ \\
\hline
\multicolumn{2}{c}{de Vaucouleurs}		& \multicolumn{2}{c}{Freeman}		& \multicolumn{3}{c}{ES}   &  \\
$0.45_{-0.03}^{+0.05}$ & $1.62_{-0.10}^{+0.15}$ & $1.59_{-0.05}^{+0.06}$ & $5.11_{-0.13}^{+0.20}$ & $3.40_{-0.44}^{+0.39}$ & $23.10_{-1.01}^{+1.85}$ & $10.45^{+2.75}_{-1.90}$ & $17.56$ \\
\hline
\multicolumn{2}{c}{de Vaucouleurs}		& \multicolumn{2}{c}{Freeman}		& \multicolumn{3}{c}{ISO}   &  \\
$0.48_{-0.04}^{+0.04}$ & $1.74_{-0.14}^{+0.14}$ & $1.21_{-0.06}^{+0.15}$ & $4.89_{-0.22}^{+0.21}$ & $17.31_{-7.84}^{+2.32}$ & $6.33_{-0.58}^{+2.37}$ & $16.58^{+1.23}_{-8.07}$ & $40.47$ \\
\hline
\multicolumn{2}{c}{de Vaucouleurs}		& \multicolumn{2}{c}{Freeman}		& \multicolumn{3}{c}{Moore}   &  \\
$0.35_{-0.02}^{+0.06}$ & $1.33_{-0.07}^{+0.21}$ & $1.00_{-0.08}^{+0.17}$ & $4.62_{-0.14}^{+0.29}$ & $3.01_{-1.54}^{+1.26}$ & $29.57_{-4.39}^{+11.61}$ & $12.39^{+1.27}_{-7.71}$ & $23.85$ \\
\hline
\multicolumn{2}{c}{de Vaucouleurs}		& \multicolumn{2}{c}{Freeman}		& \multicolumn{3}{c}{NFW} &  \\
$0.39_{-0.03}^{+0.04}$ &  $1.45_{-0.12}^{+0.14}$ & $1.08_{-0.08}^{+0.11}$ & $4.65_{-0.19}^{+0.27}$ & $3.68_{-1.27}^{+1.72}$ & $28.28_{-4.87}^{+6.85}$ & $13.62^{+9.38}_{-6.76}$ & $23.62$\\
\hline
\end{tabular}
\label{tab:second}
\end{table*}

\begin{table*}
\setlength{\tabcolsep}{.75em}
\renewcommand{\arraystretch}{1.4}
\centering
\caption{Best-fit RC parameters of M31 obtained by considering inner (ES) and main bulges (ES), disk (Freeman) and the halo models considered in this work. The halo mass has been calculated within $r=200$ kpc. The $\Delta{\rm BIC}$ value is computed with respect to the model with the ES DM halo.}
\begin{tabular}{cccccccccr}
\hline\hline
\multicolumn{2}{c}{Inner Bulge} 	& \multicolumn{2}{c}{Main Bulge}		& \multicolumn{2}{c}{Disk}		& \multicolumn{3}{c}{Halo} &  \multirow{3}{*}{$\Delta {\rm BIC}$}\\
\cline{1-9}
$M_{ib}$	&	$r_{ib}$	&	$M_{mb}$	&	$r_{mb}$	&	$M_{d}$	&	$r_{d}$&	$\rho_0$		&	$r_0$	&	$M_h$ &\\
($10^{11}$~M$_\odot$) & (kpc) & ($10^{11}$~M$_\odot$) & (kpc) & ($10^{11}$~M$_\odot$) & (kpc) & ($10^{-3}$~M$_\odot$/pc$^3$) & (kpc) & ($10^{11}$~M$_\odot$) &\\
\hline
\multicolumn{2}{c}{ES} 	& \multicolumn{2}{c}{ES}		& \multicolumn{2}{c}{Freeman}		& \multicolumn{3}{c}{Beta}  &  \\
$0.033_{-0.002}^{+0.002}$ & $0.069_{-0.003}^{+0.003}$ & $0.19_{-0.01}^{+0.02}$ & $0.44_{-0.02}^{+0.03}$ & $1.72_{-0.07}^{+0.08}$ & $5.25_{-0.15}^{+0.17}$ & $2.57_{-0.47}^{+0.45}$ & $28.09_{-1.85}^{+3.96}$ & $13.36^{+5.01}_{-5.58}$ & $4.99$\\
\hline
\multicolumn{2}{c}{ES} 	& \multicolumn{2}{c}{ES}		& \multicolumn{2}{c}{Freeman}		& \multicolumn{3}{c}{Brownstein} &  \\
$0.033_{-0.001}^{+0.001}$ & $0.069_{-0.001}^{+0.003}$ & $0.20_{-0.01}^{+0.01}$ & $0.44_{-0.01}^{+0.02}$ & $1.89_{-0.09}^{+0.03}$ & $5.53_{-0.21}^{+0.12}$ & $1.38_{-0.07}^{+0.30}$ & $33.27_{-2.68}^{+1.96}$ & $11.49^{+3.01}_{-2.35}$ & $2.07$\\
\hline
\multicolumn{2}{c}{ES} 	& \multicolumn{2}{c}{ES}		& \multicolumn{2}{c}{Freeman}		& \multicolumn{3}{c}{Burkert}  &  \\
$0.033_{-0.001}^{+0.003}$ & $0.070_{-0.004}^{+0.004}$ & $0.19_{-0.01}^{+0.02}$ & $0.43_{-0.01}^{+0.03}$ & $1.63_{-0.09}^{+0.09}$ & $5.07_{-0.14}^{+0.26}$ & $4.50_{-1.13}^{+0.81}$ & $24.87_{-1.85}^{+4.34}$ & $12.38^{+5.62}_{-3.80}$ & $7.13$ \\
\hline
\multicolumn{2}{c}{ES} 	& \multicolumn{2}{c}{ES}		& \multicolumn{2}{c}{Freeman}		& \multicolumn{3}{c}{ES} &  \\
$0.033_{-0.002}^{+0.002}$ & $0.069_{-0.003}^{+0.003}$ & $0.20_{-0.02}^{+0.01}$ & $0.45_{-0.03}^{+0.02}$ & $1.79_{-0.08}^{+0.03}$ & $5.43_{-0.23}^{+0.05}$ & $2.88_{-0.26}^{+0.41}$ & $24.59_{-1.00}^{+1.40}$ & $10.64^{+2.34}_{-1.59}$ & $0.00$ \\
\hline
\multicolumn{2}{c}{ES} 	& \multicolumn{2}{c}{ES}		& \multicolumn{2}{c}{Freeman}		& \multicolumn{3}{c}{ISO} &  \\
$0.033_{-0.001}^{+0.003}$ & $0.069_{-0.002}^{0.004}$ & $0.21_{-0.01}^{+0.01}$ & $0.45_{-0.01}^{+0.03}$ & $1.27_{-0.05}^{+0.17}$ & $5.21_{-0.33}^{+0.12}$ &  $36.08_{-12.34}^{+12.34}$ & $4.21_{-3.28}^{+3.28}$ & $15.54^{+23.81}_{-10.22}$ & $20.17$\\
\hline
\multicolumn{2}{c}{ES} 	& \multicolumn{2}{c}{ES}		& \multicolumn{2}{c}{Freeman}		& \multicolumn{3}{c}{Moore} &  \\
$0.033_{-0.002}^{+0.001}$ & $0.070_{-0.004}^{+0.001}$ & $0.19_{-0.01}^{+0.01}$ & $0.44_{-0.02}^{+0.01}$ & $1.30_{-0.04}^{+0.10}$ & $5.01_{-0.13}^{+0.15}$ & $2.54_{-0.93}^{+0.47}$ & $33.05_{-2.45}^{+8.40}$ & $13.59^{+8.54}_{-5.50}$ & $11.32$\\
\hline
\multicolumn{2}{c}{ES} 	& \multicolumn{2}{c}{ES}		& \multicolumn{2}{c}{Freeman}		& \multicolumn{3}{c}{NFW} &  \\
$0.033_{-0.001}^{+0.002}$ & $0.069_{-0.003}^{+0.004}$ & $0.18_{-0.01}^{+0.02}$ & $0.43_{-0.01}^{+0.03}$ & $1.27_{-0.09}^{+0.12}$ & $4.92_{-0.13}^{+0.28}$ & $2.49_{-0.83}^{+1.40}$ & $33.93_{-6.71}^{+7.62}$ & $13.13^{+9.95}_{-7.45}$ & $11.29$ \\
\hline
\end{tabular}
\label{tab:third}
\end{table*}

Table~\ref{tab:third} lists the best-fit results obtained from the $\sim0$--$400$~kpc RC data by splitting the bulge into inner and main components in analogy to \citet{Sofue2013}.
We fixed the exponential sphere density profile for both inner and main bulges and the Freeman density profile for the disk, whereas for the halo we explored all the DM density profiles in Eqs.~\eqref{eq:sample1}--\eqref{eq:sample6} and the exponential sphere profile.

To compare the results of Tables~\ref{tab:second}--\ref{tab:third}, we computed the difference $\Delta{\rm BIC}$ with respect to the model providing the lowest value BIC$_0$.
According to this test, it turned out that the models with ES density profiles for both inner and main bulges indeed fit better than the de Vaucouleurs profile for a single bulge, with the latter models providing $\Delta{\rm BIC}\gtrsim12$--$20$ with respect to the former ones.
In particular, in Table~\ref{tab:third} we see that the best-fit model with the lowest value BIC$_0=224.35$ is the two-bulge model with a DM halo described by the ES density profile; also the Brownstein profile represents a good fit for the DM halo, being $\Delta{\rm BIC}\approx2$.
The NFW profile, like the Moore profile, does not perform well, since $\Delta{\rm BIC}\approx11$.
In absolute, the worst fit is given by the ISO profile with $\Delta{\rm BIC}\approx20$; this is also certified by the fact that the RC of M31 is not flat in the halo region \citep[see][for details]{Sofue2015}.

Also in the case of fits with a single bulge described by the de Vaucouleurs profile (strongly excluded with respect to the fits with two bulges), among the considered DM density profiles, ES profile ($\Delta{\rm BIC}\approx18$) and Brownstein profile ($\Delta{\rm BIC}\approx19$) fit better than NFW and Moore profiles ($\Delta{\rm BIC}\approx24$), whereas the ISO profile ($\Delta{\rm BIC}\approx40$) is definitely the worst-performing one (see Table~\ref{tab:second}).

We performed also more conservative approaches involving two de Vaucouleurs profiles for inner and main bulges.
However, we obtained higher BIC and $\Delta$BIC values (not reported here for the sake of clarity and simplicity of the presentation), certifying that two ES profiles provide better a fit.
\begin{figure*}
\centering
\includegraphics[width=0.47\linewidth]{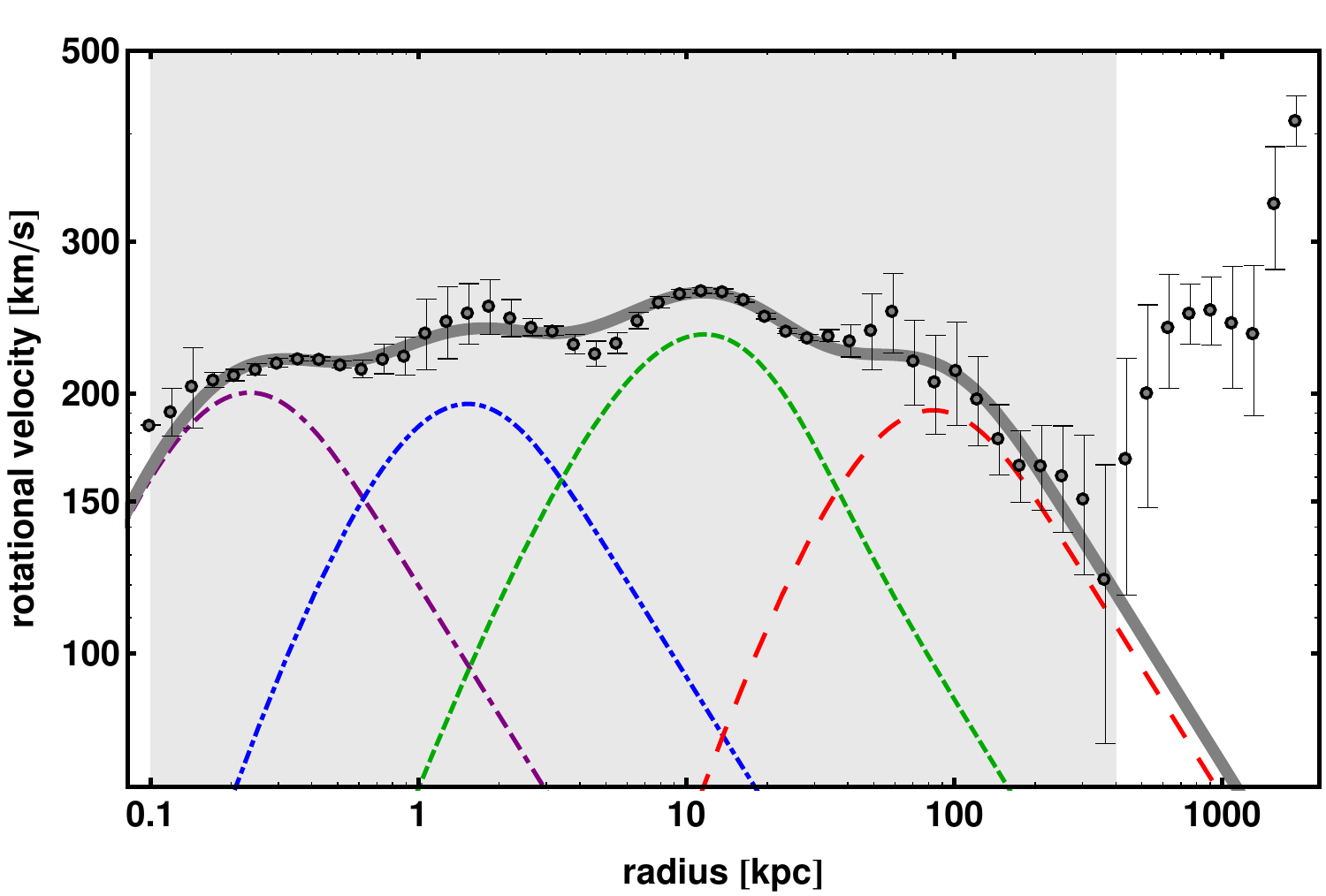}
\hfill
\includegraphics[width=0.47\linewidth]{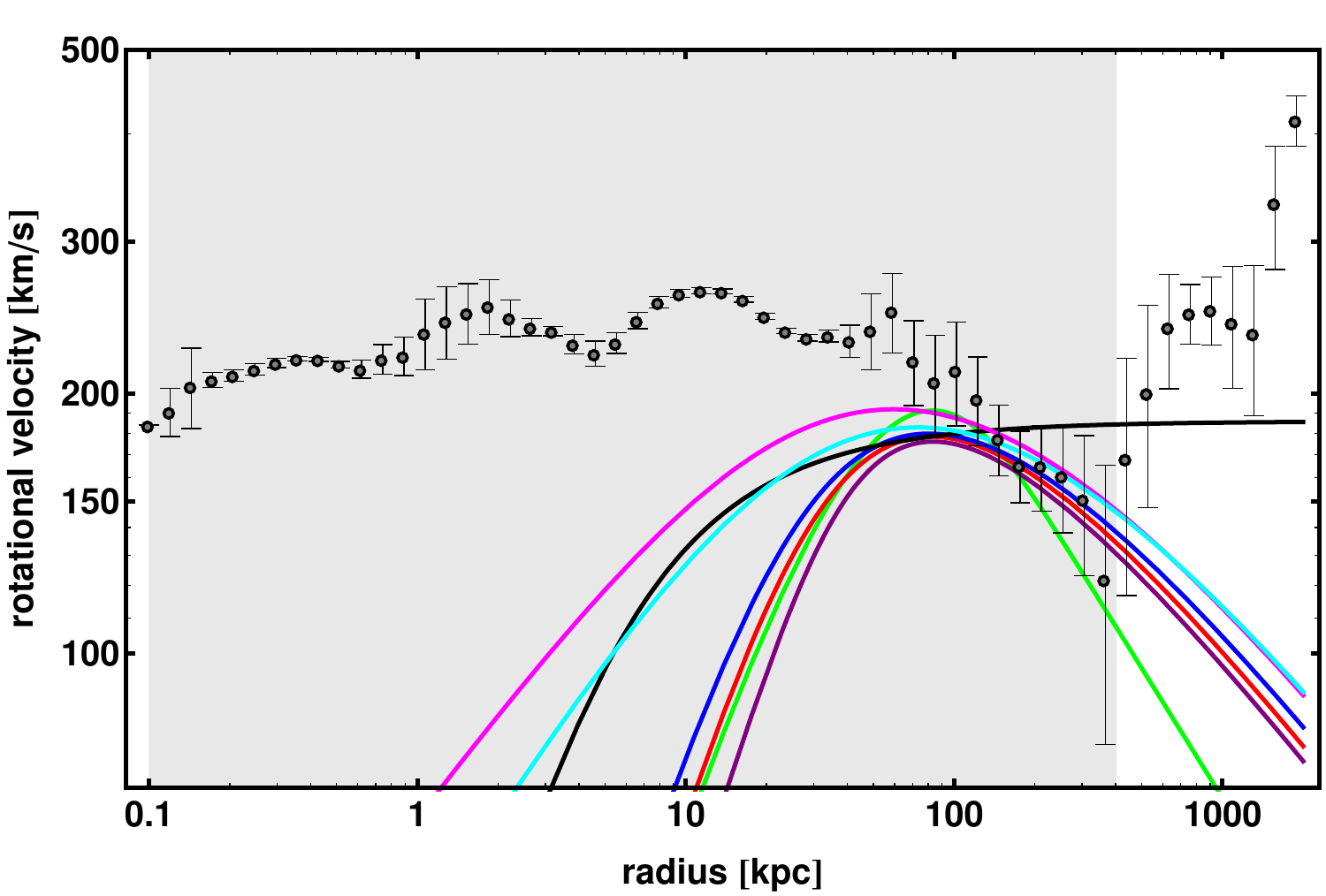}
\caption{Reconstructed RC of M31. Like in Fig.~\ref{fig:M31_Sofue}, the shaded area marks the data (black dots with error bars) considered for the fitting. Left panel: best-fit total RC (thick, gray) composed of inner bulge (ES, dash-dash-dotted purple), main bulge (ES, dot-dashed blue), disk (Freeman, dashed green) and halo (ES, long-dashed red) curves. Right panel: RC of the halo of M31 with ISO (black), Burkert (blue), Beta (red), Moore (magenta), ES (green), NFW (cyan) and Brownstein (purple) profiles, obtained from the fitting with inner and main bulges (see Table~\ref{tab:second}).}
\label{fig:M31}
\end{figure*}
\begin{figure*}
\centering
\includegraphics[width=0.47\linewidth]{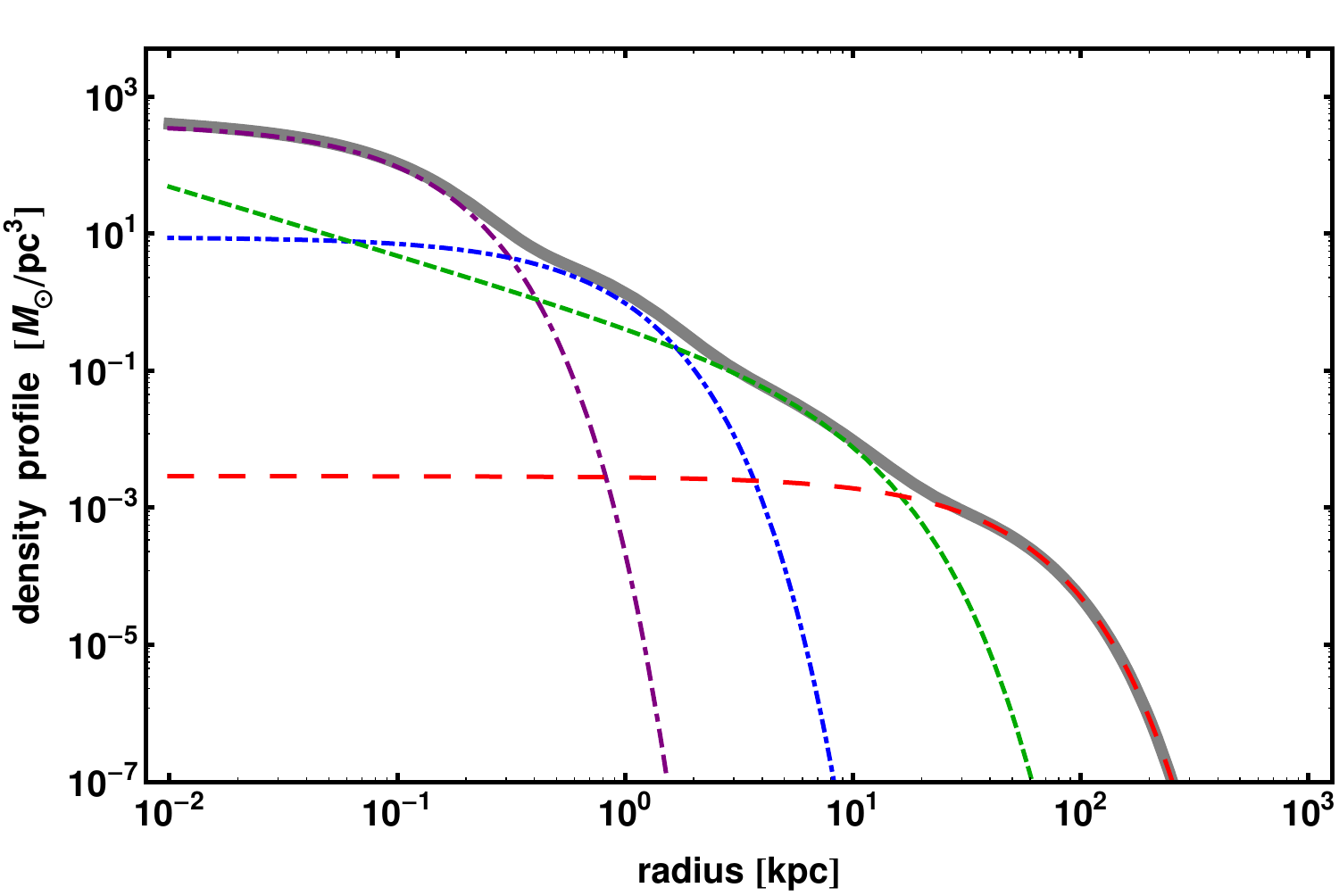}
\hfill
\includegraphics[width=0.47\linewidth]{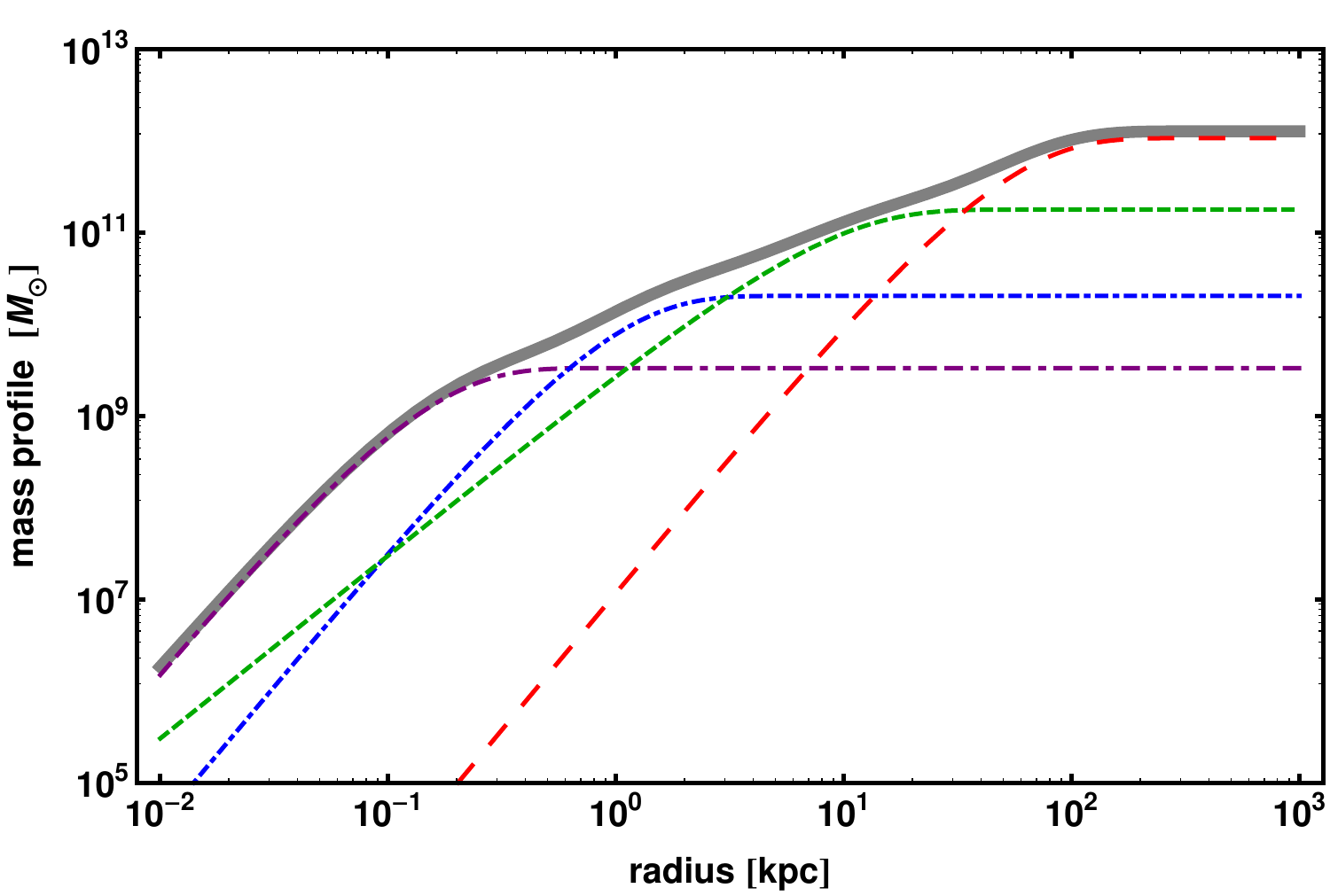}\\
\includegraphics[width=0.47\linewidth]{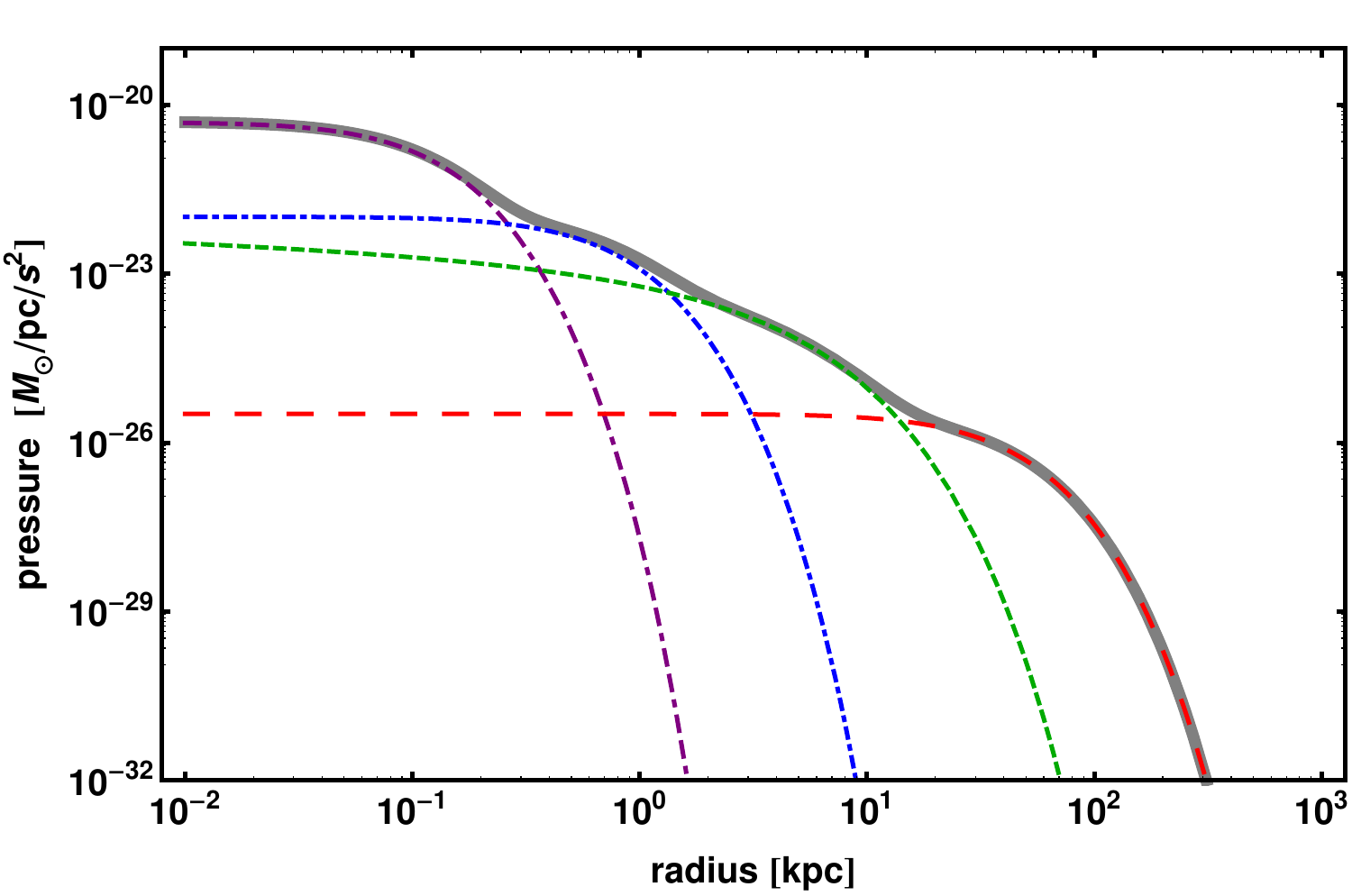}
\hfill
\includegraphics[width=0.47\linewidth]{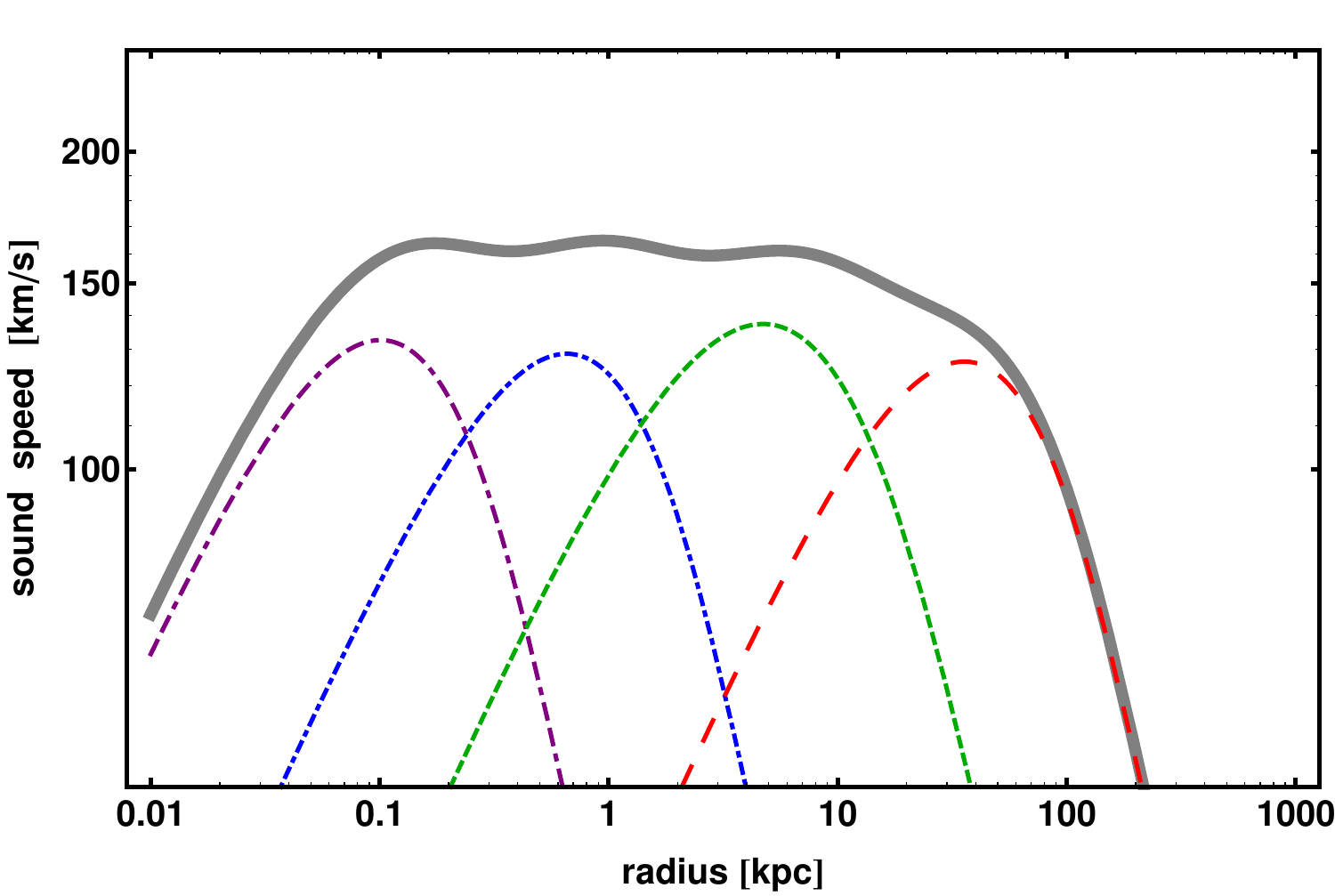}
\caption{Density (top left), mass (top right), pressure (bottom left) and speed of sound (bottom right) profiles of M31 for the best-fit combination of models (see Table~\ref{tab:third}). The net profiles and the curves for inner and main bulges, disk and DM halo have the same symbols and colors of the left panel of Fig.~\ref{fig:M31}.}
\label{fig:rho_M_P_cs}
\end{figure*}

The RC data with error bars of the M31 galaxy  is illustrated in the left panel of Fig.~\ref{fig:M31}.
Superposed on the observational data, the best-fit curves are also displayed: the inner and main bulges modeled with two exponential density profiles, the disk with the Freeman profile, and the halo region with the exponential density model (see values in Table~\ref{tab:third}).
On the right panel of Fig.~\ref{fig:M31}, we focused on the halo sector of the RC of M31, showing all the DM profiles considered in this work. It is worth noticing that, unlike DM profiles for halos, there are not so many alternatives for the inner parts of galaxies.

Fig.~\ref{fig:rho_M_P_cs} displays the density $\rho(r)$ (top left panel), the mass $M(r)$ (top right panel), the pressure $P(r)$ (bottom left panel) and the speed of sound $c_s=\sqrt{\partial P/\partial\rho}$ (bottom right panel) radial distributions obtained from the combination of models that best-fit inner bulge, main bulge, disk and DM halo of M31 (see Table~\ref{tab:third}). As expected the main contribution to the mass of the galaxy is given by the presence of DM in the halo.
Following \citet{2015MNRAS.449..403B}, the pressure  in the whole galaxy was assumed to be non-zero and has been estimated from the Newtonian hydrostatic equilibrium equations
\begin{subequations}
\begin{align}
\label{eq:massbalance}
\frac{d M(r)}{dr}&=4\pi r^2 \rho(r)\,,\\
\label{eq:pressbalance}
\frac{d P(r)}{dr}&=-\rho(r)\frac{GM(r)}{r^2}\,.
\end{align}
\end{subequations}
The speed of sound, obtained from the hydrostatic equilibrium equation, is given by the expression
\be
c_s(r)=V(r)\sqrt{-\frac{d\ln r}{d\ln \rho(r)}}\,,
\ee
from which it is clear its correlation with the RC velocity.

\subsection{Comparison with the Milky Way galaxy}

The MW is a spiral galaxy with a structure similar to M31 and so it appears natural and needful a direct comparison between the two galaxies.
The RC data of the MW has been updated in \citet{Sofue2015}, taking into account the linear speed of the Sun with respect to the galactic center as $V_\odot=238$~km/s instead of $V_\odot=200$~km/s \citep{Sofue2013}.

\begin{figure}
\centering
\includegraphics[width=0.97\linewidth]{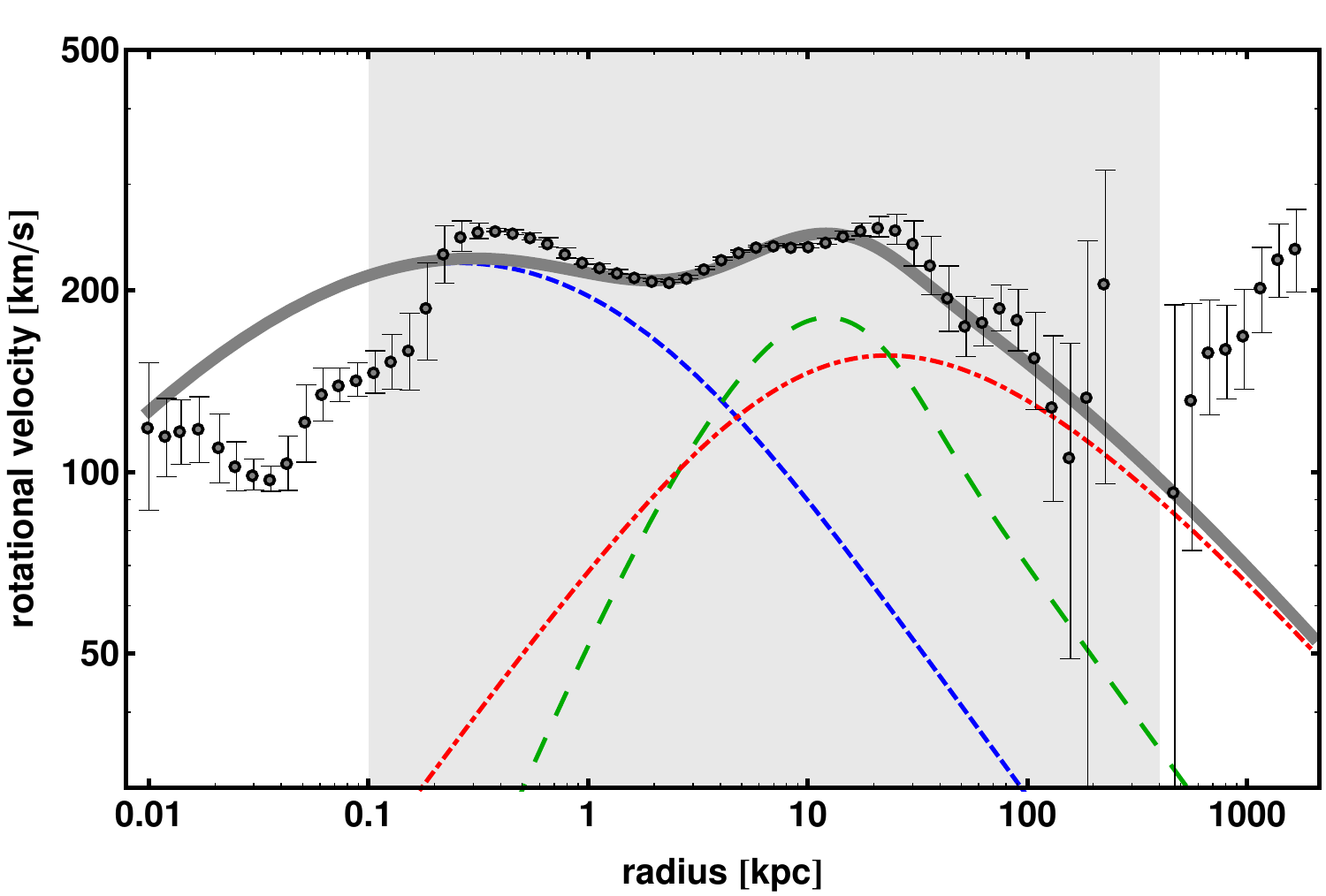}
\caption{Best-fit RC of the MW (thick, gray) composed of bulge (de Vaucouleurs, dot-dashed blue), disk (Freeman, dashed green) and halo (NFW, long-dashed red) components. The shaded area marks the data (black dots with error bars) considered for the fitting. Reproduced from \citet{Sofue2015}.}
\label{fig:MW_Sofue}
\end{figure}

Fig.~\ref{fig:MW_Sofue} and Tables~\ref{tab:MW_first} reproduce the results of the RC analysis of the MW obtained by \citet{Sofue2015}. In analogy to M31, the bulge is modelled by the de Vaucouleurs profile, the disk is approximated by the exponential disk model and the halo was described by the NFW model. The fitting was performed within the range $0$--$20$ kpc for the bulge, $0$--$40$ kpc for the disk, and $0$--$385$~kpc for the halo \citep{Sofue2015}.

The MW RC data and corresponding detailed analyses, invoking the ES density profile for both inner and main bulges, the Freeman model for the disk and the NFW profile for the halo region, have been presented in \citet{Sofue2013}. The extension of the analyses by \citet{Sofue2013} was given in \citet{2021MNRAS.508.1543B}, where the ES density profile has also been used to interpret the RC data in the disk and halo of the galaxy along with other widely exploited profiles. However, in \citet{2021MNRAS.508.1543B} in order to be in consistent with observations, the scale radius of the halo has been fixed to $12$~kpc, which substantially facilitated the analyses. However, here we leave free the scale radius and infer it from the fit, which we presume is more reasonable than fixing it in advance.

Table~\ref{tab:MW_one_bulge} lists the best-fit results obtained from the $\sim0.1$--$400$~kpc RC by considering a single bulge \citet{Sofue2015}. Like for the case of M31, we here fixed the de Vaucouleurs profile for the bulge, the Freeman profile for the disk, whereas for the halo we explored the profiles in Eqs.~\eqref{eq:sample1}--\eqref{eq:sample6} and the ES profile. Again, the last combination of models listed in Table~\ref{tab:MW_one_bulge} represents a direct comparison, though performed with a different methodology (see Sec.~\ref{sec:mass}), with the results of \citet{Sofue2015} summarized Table~\ref{tab:MW_first}.

Table~\ref{tab:MW_two_bulges} lists the best-fit results obtained from the $\sim0.01$--$400$~kpc RC data by splitting the bulge into inner and main components in analogy to \citet{Sofue2013}.
We fixed the exponential sphere density profile for both inner and main bulges and the Freeman density profile for the disk, whereas for the halo we explored all the DM density profiles in Eqs.~\eqref{eq:sample1}--\eqref{eq:sample6} and the exponential sphere profile.

The comparison between the results of Tables~\ref{tab:MW_one_bulge}--\ref{tab:MW_two_bulges} is not straightforward. The reason is that to reproduce the results of \citet{Sofue2015} summarized in Table~\ref{tab:MW_first} and to perform the analysis shown in Table~\ref{tab:MW_one_bulge}, the fits of the one bulge cases focused on the range $\sim0.1$--$400$~kpc, whereas the fits of the two bulge cases focused on the range $\sim0.01$--$400$~kpc to perform better modeling of the inner bulge.
However, taking in mind the above caveat, we performed the BIC tests for both cases and computed the difference $\Delta{\rm BIC}$ with respect to the model providing the lowest value BIC$_0$.
According to this test, it turned out that the models with ES density profiles for both inner and main bulges indeed fit better than the de Vaucouleurs profile for a single bulge, with the latter models providing $\Delta{\rm BIC}\gtrsim100$ with respect to the former ones.
In particular, in Table~\ref{tab:MW_two_bulges} we see that the best-fit model with the lowest value BIC$_0=134.16$ is the two-bulge model with a DM halo described by the ES density profile.
The NFW profile, like the Moore profile, does not perform well, since $\Delta{\rm BIC}\approx16$.
In absolute, the worst fit is given by the ISO profile with $\Delta{\rm BIC}\approx58$, as certified by the fact that the RC of the MW is not flat in the halo region (see Figure~\ref{fig:MW_Sofue}).

Also among the fits with a single bulge described by the de Vaucouleurs profile (strongly excluded with respect to the fits with inner and main bulges), the DM ES profile ($\Delta{\rm BIC}\approx160$) and Brownstein profile ($\Delta{\rm BIC}\approx166$) fit better than NFW profile ($\Delta{\rm BIC}\gtrsim191$), whereas the ISO profile ($\Delta{\rm BIC}\approx197$) is definitely the worst-performing one (see Table~\ref{tab:MW_one_bulge}).

Finally, fits with two de Vaucouleurs profiles for inner and main bulges lead to higher BIC and $\Delta$BIC values. Again, these analyses have not been shown here to keep the presentation clear and simple.

\begin{figure*}
\centering
\includegraphics[width=0.47\linewidth]{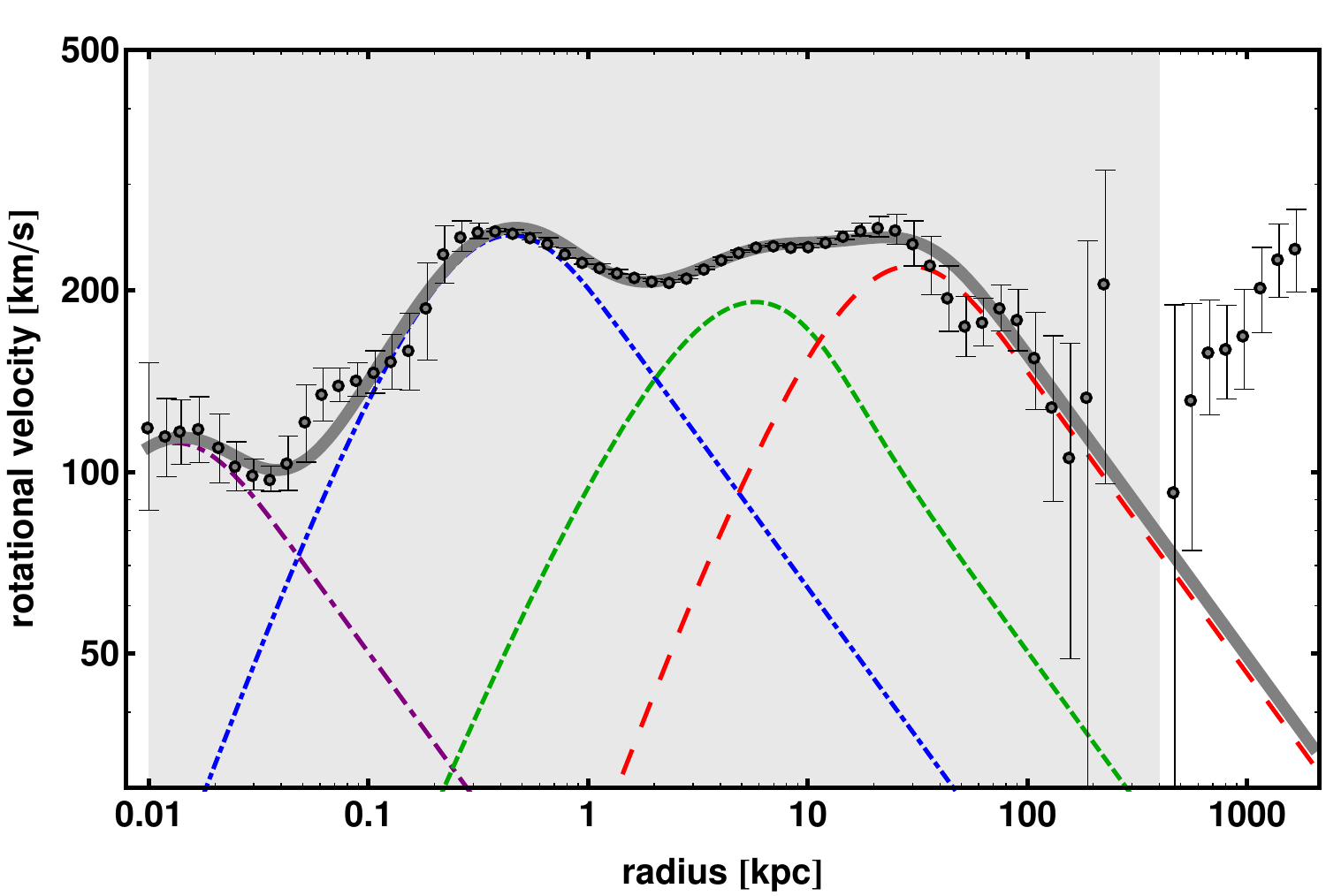}
\hfill
\includegraphics[width=0.47\linewidth]{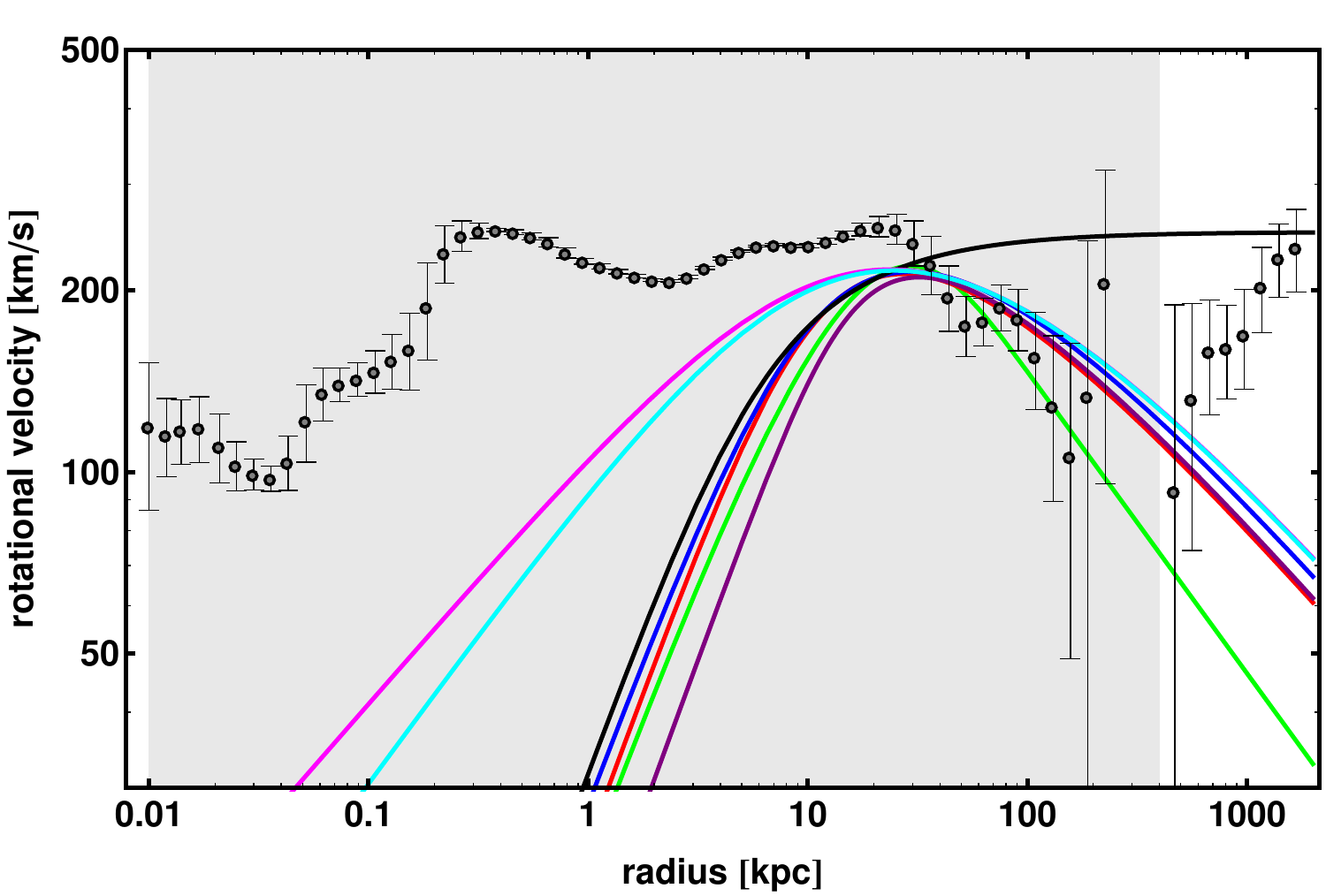}
\caption{Reconstructed RC of the MW. Left panel: best-fit total RC composed of inner and main bulges, disk and ES profile for the halo. Right panel: RC of the halo of the MW with Beta, Brownstein, Burkert, ES, ISO, Moore and NFW profiles.
The net profiles and the curves for inner and main bulges, disk and DM halo have the same symbols and colors of the left panel of Fig.~\ref{fig:M31}. The best-fit values are taken from Table~\ref{tab:MW_two_bulges}.}
\label{fig:image1ab}
\end{figure*}

The left panel of Fig.~\ref{fig:image1ab} portrays the RC data with error bars of the MW galaxy with the best-fitting curve composed of the inner and main bulges modeled with two ES profiles, the disk with the Freeman profile, and the halo region with the ES model (see values in Table~\ref{tab:MW_two_bulges}).
The right panel of Fig.~\ref{fig:image1ab} displays the halo RC part with the DM profiles considered in this work (see the values in Table~\ref{tab:MW_two_bulges}) .

\begin{figure*}
\centering
\includegraphics[width=0.47\linewidth]{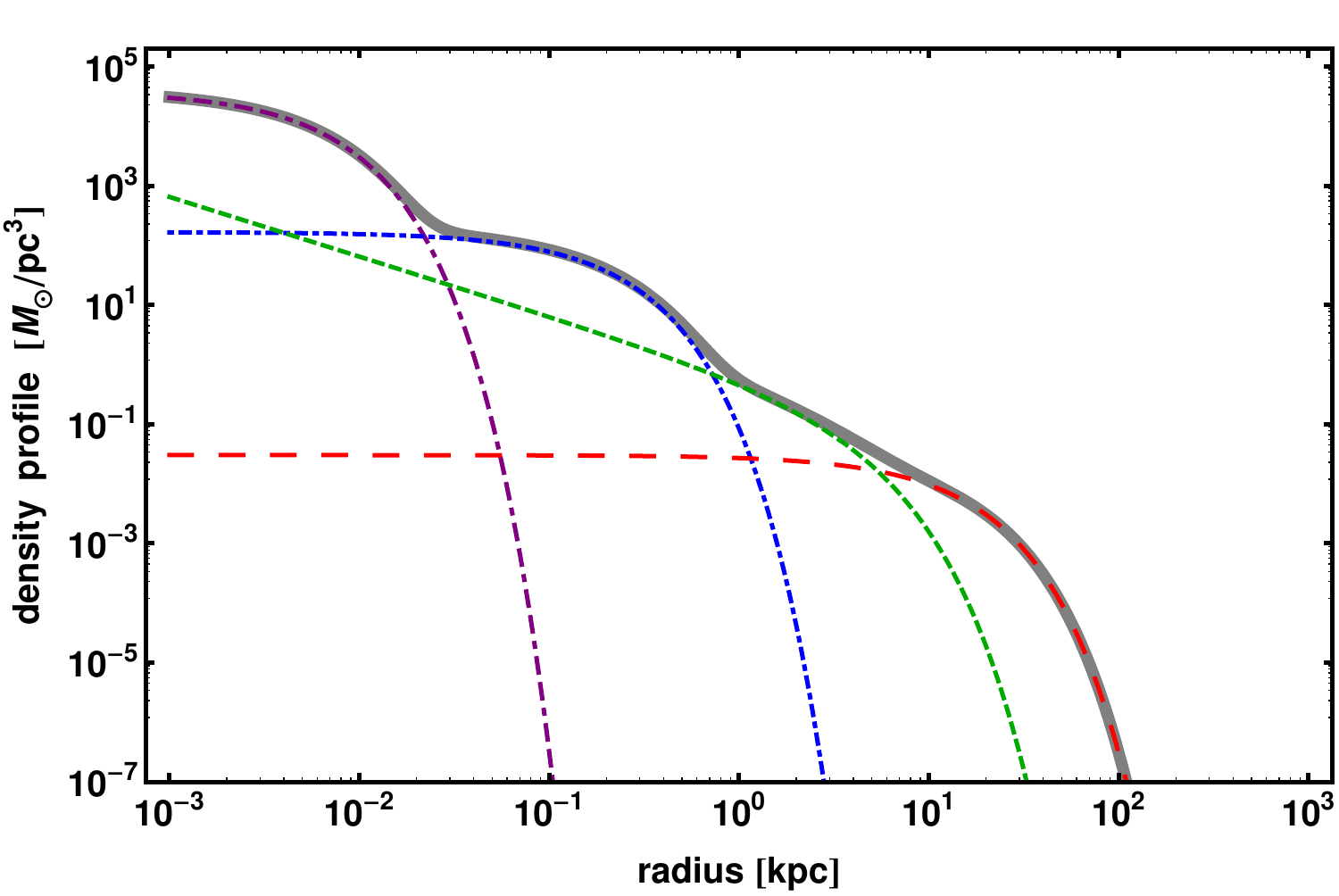}
\hfill
\includegraphics[width=0.47\linewidth]{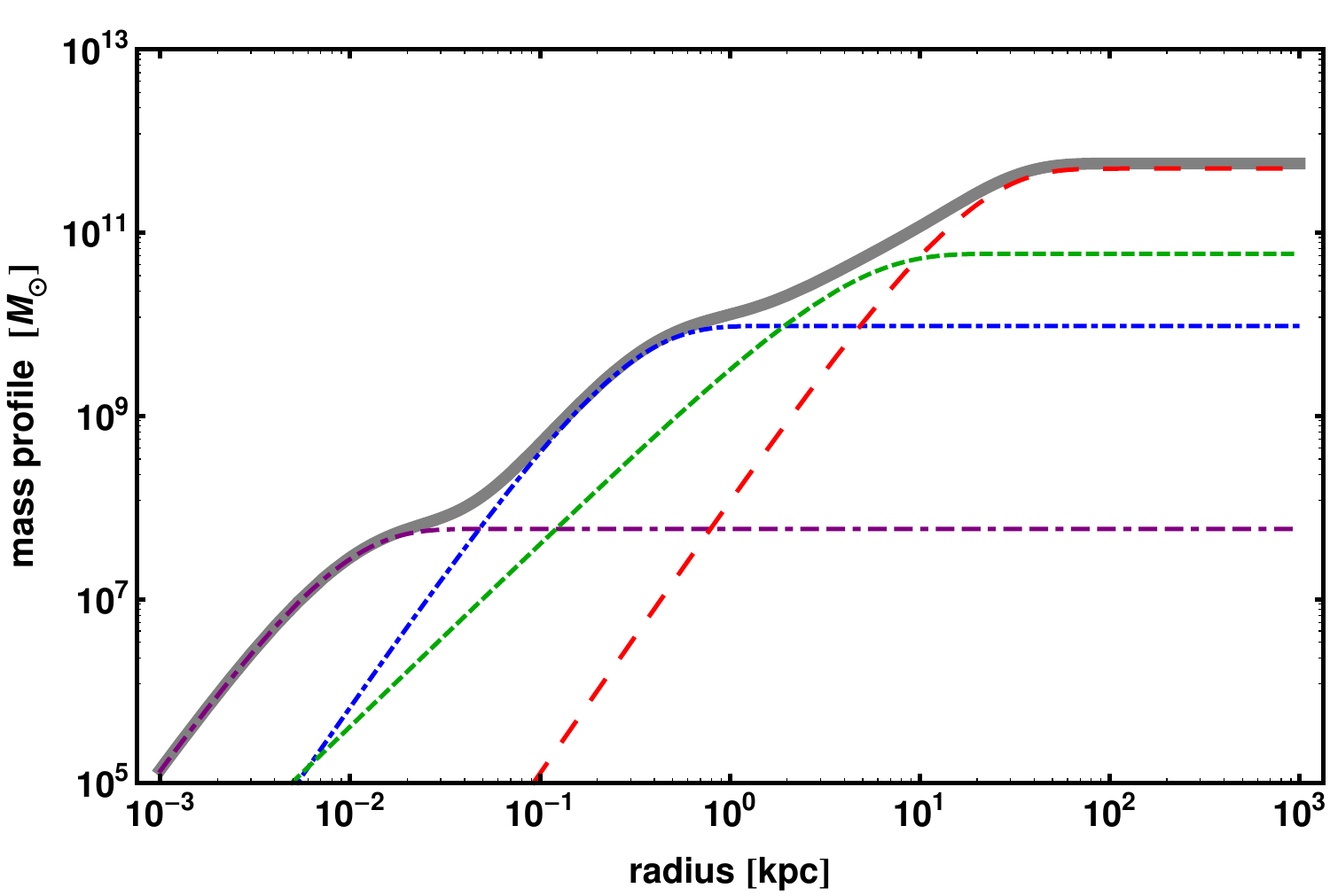}\\
\includegraphics[width=0.47\linewidth]{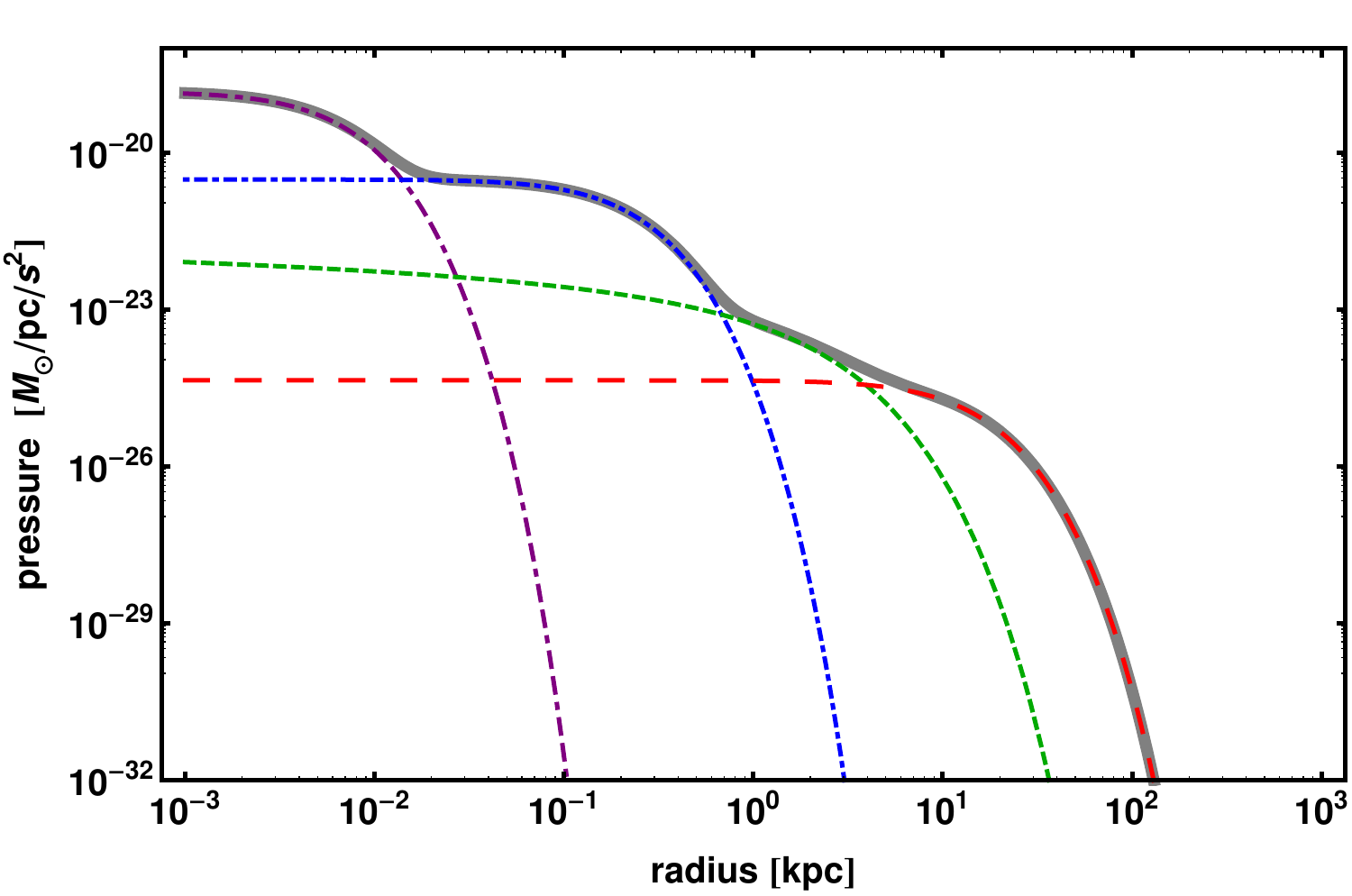}
\hfill
\includegraphics[width=0.47\linewidth]{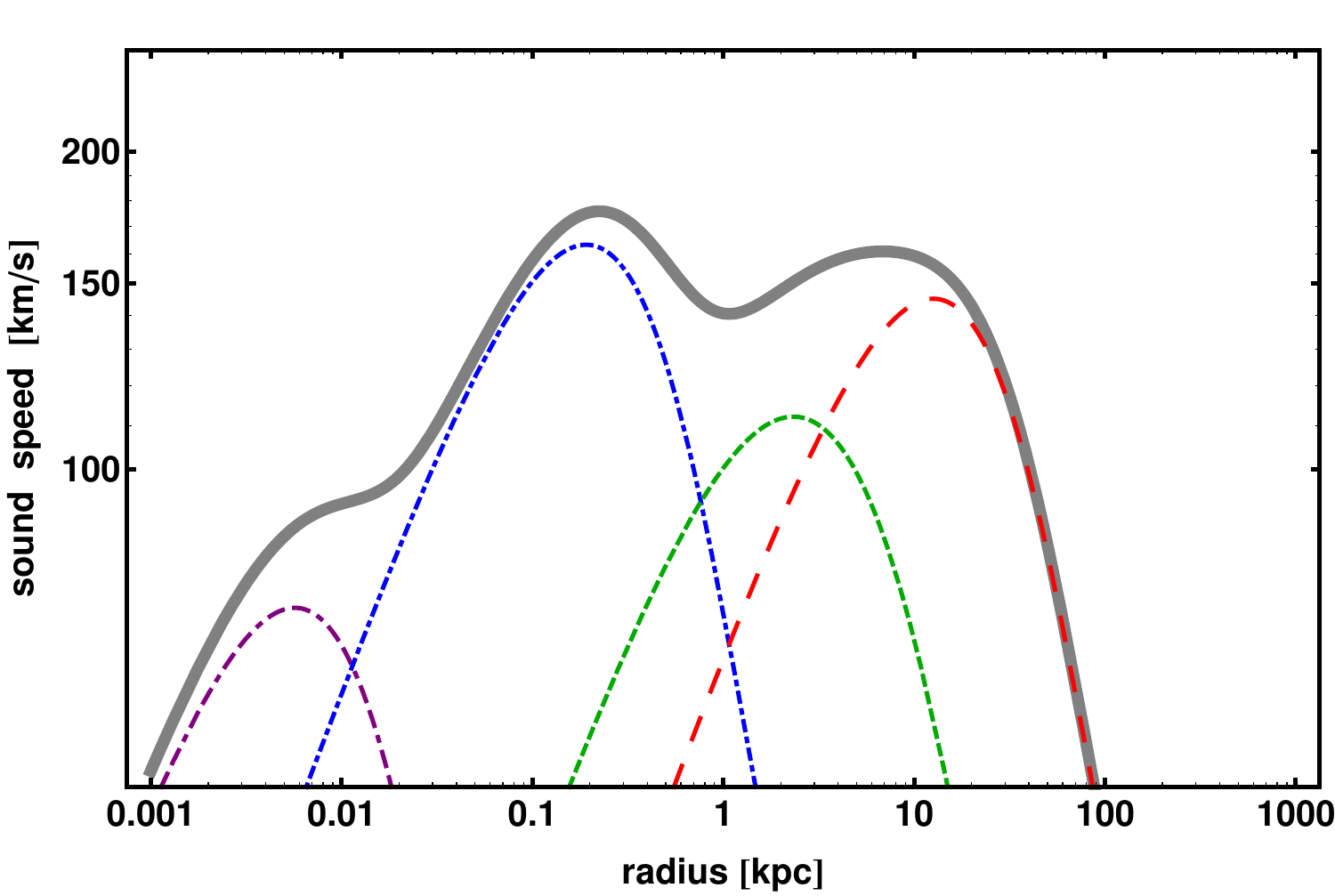}
\caption{Density (top left), mass (top right), pressure (bottom left) and speed of sound (bottom right) profiles of the MW for the best-fit combination of models (see Table~\ref{tab:MW_two_bulges}). The net profiles and the curves for inner and main bulges, disk and DM halo have the same symbols and colors of the left panel of Fig.~\ref{fig:image1ab}.}
\label{fig:rho_M_P_cs_MW}
\end{figure*}

Fig.~\ref{fig:rho_M_P_cs_MW} displays the density $\rho(r)$ (top left panel), the mass $M(r)$ (top right panel), the pressure $P(r)$ (bottom left panel) and the speed of sound $c_s$ (bottom right panel) radial distributions obtained from the combination of models best-fitting the RC of the MW (see Fig.~\ref{fig:image1ab} and  Table~\ref{tab:MW_two_bulges}).
Similar results have been obtained for the MW in \citet{2021MNRAS.508.1543B}.
\begin{table}
\centering
\caption{The best-fit parameters of the MW from \citet{Sofue2015}. The halo model is the NFW profile and its mass has been calculated within $r=200$ kpc.}
\label{tab:MW_first}
\begin{tabular}{cccc}
\hline
Parts   & Total mass              & Scale radius          & Central density\\
        &  ($10^{11}M_{\odot}$)   &  (kpc)
        &  ($10^{-3} M_{\odot}$/pc$^3$) \\
\hline
Bulge        & $0.25\pm 0.02$         & $0.87\pm 0.07$   & --- \\

Disk         & $1.12\pm 0.40$          & $5.73\pm 1.23$   & ---  \\

Halo         & $5.7\pm 5.1$.        & $10.7\pm 2.9$      & $18.2\pm 7.4$  \\
 \hline
\end{tabular}
\end{table}

\begin{table*}
\setlength{\tabcolsep}{1.6em}
\renewcommand{\arraystretch}{1.4}
\centering
\caption{Best-fit RC parameters of the MW  obtained by considering one bulge (de Vaucouleurs), disk (Freeman) and the halo models considered in this work. The halo mass has been calculated within $r=200$~kpc. The $\Delta{\rm BIC}$ value is computed with respect to the minimum BIC in Table~\ref{tab:MW_two_bulges}.}
\begin{tabular}{cccccccr}
\hline\hline
\multicolumn{2}{c}{Bulge}		& \multicolumn{2}{c}{Disk}		& \multicolumn{3}{c}{Halo}  &  \multirow{3}{*}{$\Delta {\rm BIC}$}\\
\cline{1-7}
$M_{b}$	&	$r_{b}$	&	$M_{d}$	&	$r_{d}$&	$\rho_0$		&	$r_0$	&	$M_h$ &\\
($10^{11}$~M$_\odot$) & (kpc) & ($10^{11}$~M$_\odot$) & (kpc) & ($10^{-3}$~M$_\odot$/pc$^3$) & (kpc) & ($10^{11}$~M$_\odot$) &\\
\hline
\multicolumn{2}{c}{de Vaucouleurs}		& \multicolumn{2}{c}{Freeman}		& \multicolumn{3}{c}{Beta} &  \\
$0.20_{-0.01}^{+0.01}$ & $0.49_{-0.03}^{+0.03}$ & $0.75_{-0.07}^{+0.09}$ & $3.38_{-0.17}^{+0.19}$ & $15.02_{-3.72}^{+3.74}$ & $12.82_{-1.43}^{+1.89}$ & $9.72^{+4.43}_{-3.70}$  & $172.57$\\
\hline
\multicolumn{2}{c}{de Vaucouleurs}      & \multicolumn{2}{c}{Freeman}		& \multicolumn{3}{c}{Brownstein} &  \\
$0.20_{-0.01}^{+0.01}$ & $0.49_{-0.02}^{+0.02}$ & $0.89_{-0.04}^{+0.06}$ & $3.61_{-0.12}^{+0.14}$ & $7.84_{-1.24}^{+0.83}$ & $15.45_{-0.67}^{+1.40}$ & $9.32^{+2.41}_{-1.81}$ & $165.56$ \\
\hline
\multicolumn{2}{c}{de Vaucouleurs}      & \multicolumn{2}{c}{Freeman}		& \multicolumn{3}{c}{Burkert} &  \\
$0.20_{-0.01}^{+0.01}$ & $0.50_{-0.04}^{+0.01}$ & $0.50_{-0.02}^{+0.15}$ & $2.95_{-0.09}^{+0.29}$ & $47.65_{-19.39}^{+1.85}$ & $8.55_{-0.15}^{+2.35}$ & $9.02^{+6.45}_{-3.69}$ & $172.53$\\
\hline
\multicolumn{2}{c}{de Vaucouleurs}		& \multicolumn{2}{c}{Freeman}		& \multicolumn{3}{c}{ES} &  \\
$0.20_{-0.01}^{+0.01}$ & $0.48_{-0.03}^{+0.02}$ & $0.74_{-0.05}^{+0.06}$ & $3.37_{-0.12}^{+0.14}$ & $20.33_{-3.03}^{+3.03}$ & $9.97_{-0.60}^{+0.80}$ & $5.07^{+1.43}_{-1.18}$ & $160.30$\\
\hline
\multicolumn{2}{c}{de Vaucouleurs}		& \multicolumn{2}{c}{Freeman}		& \multicolumn{3}{c}{ISO} &  \\
$0.21_{-0.01}^{+0.01}$ & $0.53_{-0.02}^{+0.03}$ & $0.80_{-0.02}^{+0.03}$ & $3.62_{-0.07}^{+0.12}$ & $19.30_{-1.59}^{+0.68}$ & $6.81_{-0.41}^{+0.24}$ & $21.30^{+1.66}_{-3.07}$ & $197.73$\\
\hline
\multicolumn{2}{c}{de Vaucouleurs}		& \multicolumn{2}{c}{Freeman}		& \multicolumn{3}{c}{Moore} &  \\
$0.18_{-0.01}^{+0.01}$ & $0.43_{-0.02}^{+0.03}$ & $0.67_{-0.07}^{+0.07}$ & $3.89_{-0.11}^{+0.13}$ & $4.33_{-0.86}^{+1.71}$ & $25.22_{-3.49}^{+3.11}$ & $12.15^{+6.02}_{-4.73}$ & $196.29$\\
\hline
\multicolumn{2}{c}{de Vaucouleurs}		& \multicolumn{2}{c}{Freeman}		& \multicolumn{3}{c}{NFW} & \\
$0.18_{-0.01}^{+0.01}$ & $0.44_{-0.03}^{+0.03}$ & $0.62_{-0.08}^{+0.09}$ & $3.77_{-0.20}^{+0.11}$ & $9.73_{-3.68}^{+2.70}$ & $18.05_{-1.94}^{+5.31}$ & $12.00^{+8.78}_{-5.23}$ & $191.10$ \\
\hline
\end{tabular}
\label{tab:MW_one_bulge}
\end{table*}

\begin{table*}
\setlength{\tabcolsep}{.5em}
\renewcommand{\arraystretch}{1.4}
\centering
\caption{Best-fit RC parameters of the MW obtained by considering inner (ES) and main bulges (ES), disk (Freeman) and the halo models considered in this work. The halo mass has been calculated within $r=200$ kpc. The $\Delta{\rm BIC}$ value is computed with respect to the model with the ES DM halo.}
\begin{tabular}{cccccccccr}
\hline\hline
\multicolumn{2}{c}{Inner Bulge} 	& \multicolumn{2}{c}{Main Bulge}		& \multicolumn{2}{c}{Disk}		& \multicolumn{3}{c}{Halo}  &  \multirow{3}{*}{$\Delta {\rm BIC}$}\\
\cline{1-9}
$M_{ib}$	&	$r_{ib}$	&	$M_{mb}$	&	$r_{mb}$	&	$M_{d}$	&	$r_{d}$&	$\rho_0$		&	$r_0$	&	$M_h$ &\\
($10^{11}$~M$_\odot$) & (kpc) & ($10^{11}$~M$_\odot$) & (kpc) & ($10^{11}$~M$_\odot$) & (kpc) & ($10^{-3}$~M$_\odot$/pc$^3$) & (kpc) & ($10^{11}$~M$_\odot$) &\\
\hline
\multicolumn{2}{c}{ES} 	& \multicolumn{2}{c}{ES}		& \multicolumn{2}{c}{Freeman}		& \multicolumn{3}{c}{Beta} &   \\
$0.00056_{-0.00003}^{+0.00006}$ & $0.0034_{-0.0006}^{+0.0009}$ & $0.094_{-0.001}^{+0.001}$ & $0.130_{-0.002}^{+0.002}$ & $0.48_{-0.03}^{+0.08}$ & $2.41_{-0.06}^{+0.18}$ & $33.24_{-8.83}^{+4.53}$ & $9.33_{-0.57}^{+1.69}$ & $9.37^{+4.67}_{-2.91}$ & $12.59$\\
\hline
\multicolumn{2}{c}{ES} 	& \multicolumn{2}{c}{ES}		& \multicolumn{2}{c}{Freeman}		& \multicolumn{3}{c}{Brownstein} &  \\
$0.00059_{-0.00005}^{+0.00003}$ & $0.0040_{-0.0009}^{+0.0004}$ & $0.096_{-0.002}^{+0.001}$ & $0.132_{-0.003}^{+0.001}$ & $0.68_{-0.06}^{+0.02}$ & $2.82_{-0.13}^{+0.05}$ & $13.28_{-0.70}^{+3.00}$ & $12.82_{-1.40}^{+0.34}$ & $9.66^{+2.28}_{-2.82}$ & $8.82$\\
\hline
\multicolumn{2}{c}{ES} 	& \multicolumn{2}{c}{ES}		& \multicolumn{2}{c}{Freeman}		& \multicolumn{3}{c}{Burkert} &  \\
$0.00058_{-0.00004}^{+0.00004}$ & $0.00380_{-0.0008}^{+0.0005}$ & $0.094_{-0.001}^{+0.002}$ & $0.130_{-0.002}^{+0.003}$ & $0.47_{-0.02}^{+0.02}$ & $2.43_{-0.05}^{+0.05}$ & $47.65_{-3.66}^{+2.20}$ & $9.12_{-0.21}^{+0.34}$ & $10.67^{+1.14}_{-1.04}$ & $14.49$\\
\hline
\multicolumn{2}{c}{ES} 	& \multicolumn{2}{c}{ES}		& \multicolumn{2}{c}{Freeman}		& \multicolumn{3}{c}{ES} &  \\
$0.00059_{-0.00004}^{+0.00005}$ & $0.0039_{-0.0010}^{+0.0004}$ & $0.096_{-0.003}^{+0.001}$ & $0.132_{-0.003}^{+0.001}$ & $0.59_{-0.04}^{+0.02}$ & $2.68_{-0.13}^{+0.04}$ & $30.37_{-2.76}^{+4.33}$ & $8.69_{-0.54}^{+0.47}$ & $5.01^{+1.09}_{-1.04}$ & $0.00$\\
\hline
\multicolumn{2}{c}{ES} 	& \multicolumn{2}{c}{ES}		& \multicolumn{2}{c}{Freeman}		& \multicolumn{3}{c}{ISO} & \\
$0.00063_{-0.00010}^{+0.00002}$ & $0.0043_{-0.0011}^{+0.0005}$ & $0.098_{-0.001}^{+0.00}$ & $0.133_{-0.002}^{+0.001}$ & $0.44_{-0.01}^{+0.01}$ & $2.47_{-0.05}^{+0.01}$ & $57.04_{-0.53}^{+2.45}$ & $4.50_{-0.15}^{+0.03}$ & $28.04^{+1.26}_{-1.81}$ & $58.22$\\
\hline
\multicolumn{2}{c}{ES} 	& \multicolumn{2}{c}{ES}		& \multicolumn{2}{c}{Freeman}		& \multicolumn{3}{c}{Moore} &  \\
$0.00052_{-0.00006}^{+0.00003}$ & $0.0027_{-0.0001}^{+0.0012}$ & $0.083_{-0.003}^{+0.001}$ & $0.122_{-0.004}^{+0.001}$ & $0.28_{-0.05}^{+0.05}$ & $3.03_{-0.14}^{+0.02}$ & $20.35_{-3.50}^{+3.87}$ & $13.16_{-0.88}^{+1.02}$ & $11.35^{+3.1}_{-2.74}$ & $17.72$\\
\hline
\multicolumn{2}{c}{ES} 	& \multicolumn{2}{c}{ES}		& \multicolumn{2}{c}{Freeman}		& \multicolumn{3}{c}{NFW} &   \\
$0.00056_{-0.00008}^{+0.00001}$ & $0.0036_{-0.0011}^{+0.0003}$ & $0.087_{-0.004}^{+0.001}$ & $0.124_{-0.004}^{+0.002}$ & $0.29_{-0.09}^{+0.01}$ & $2.79_{-0.16}^{+0.06}$ & $30.20_{-2.57}^{+13.33}$ & $11.49_{-1.75}^{+0.53}$ & $11.31^{+5.17}_{-4.50}$ & $16.31$ \\
\hline
\end{tabular}
\label{tab:MW_two_bulges}
\end{table*}


\begin{table}
\setlength{\tabcolsep}{.65em}
\renewcommand{\arraystretch}{1.4}
\centering
\caption{Comparison of the best-fit parameters for M31 and the MW.}
\label{tab:comparison}
\begin{tabular}{llcc}
\hline
Component & Parameter & M31  & Milky Way \\
\hline
\hline
Inner bulge & $r_{ib}$ (kpc)                                & $0.069_{-0.003}^{+0.003}$
            & $0.0039_{-0.0010}^{+0.0004}$ \\
            & $M_{ib}$ ($10^{11}M_\odot$)
            & $0.033_{-0.002}^{+0.002}$
            & $0.00059_{-0.00004}^{+0.00005}$ \\
Main bulge  & $r_{mb}$ (kpc)                                & $0.45_{-0.03}^{+0.02}$
            & $0.132_{-0.003}^{+0.001}$ \\
            & $M_{mb}$ ($10^{11}M_\odot$)                   & $0.20_{-0.02}^{+0.01}$
            & $0.096_{-0.003}^{+0.001}$ \\
Disk        & $r_{d}$ (kpc)                                 & $5.43_{-0.23}^{+0.05}$
            & $2.68_{-0.13}^{+0.04}$ \\
            & $M_d$ ($10^{11}M_\odot$)                      & $1.79_{-0.08}^{+0.03}$
            & $0.59_{-0.04}^{+0.02}$ \\
Halo        & $r_0$ (kpc)                                   & $24.59_{-1.00}^{+1.40}$
            & $8.69_{-0.54}^{+0.47}$ \\
            & $\rho_0$ ($10^{-3}M_\odot/$pc$^3$)            & $2.88_{-0.26}^{+0.41}$
            & $30.37_{-2.76}^{+4.33}$ \\
            & $M_{h}$ ($10^{11}M_\odot$)
            & $10.64^{+2.34}_{-1.59}$
            & $5.01^{+1.09}_{-1.04}$ \\
Total mass  & $M_{tot}$ ($10^{11}M_\odot$)
            & $12.67^{+2.34}_{-1.59}$
            & $5.70^{+1.09}_{-1.04}$ \\
\hline
\end{tabular}
\end{table}

\section{Final outlooks and perspectives}\label{sec:concl}

In this work, the rotation curve of the Andromeda galaxy was analyzed from the galactic center to the halo region. Following \cite{Sofue2013}, the exponential sphere profile was involved in describing the observational data in the bulge and disk of the galaxy. Moreover, the bulge was decomposed into inner and main parts in analogy to \cite{Sofue2013}, where the same approach was applied to the Milky Way galaxy. The NFW, Burkert, Moore, Isothermical, Beta and Brownstein along with the exponential density profiles were used in the halo region to analyze the rotation curve for the Andromeda Galaxy.

The Bayesian information criterion (BIC)~\citep{Yunis} was used to find the model that showed the best results for a given galaxy out of six models. Among all of the models the exponential density and Brownstein profiles showed good results for the Andromeda galaxy in the halo region. The exponential sphere profile was better for the bulge with respect to the de Vaucouleurs model.

Table~\ref{tab:first}, which we retrieved from \cite{Sofue2015}, is a reference and starting point for our analyses. Knowing what we approximately expect to obtain, we extended the results of Table~\ref{tab:first} in Table~\ref{tab:second} showing that the NFW profile is good, but not the best profile among the existing alternatives. Furthermore, we continued our analyses in Table~\ref{tab:third}. Following \cite{Sofue2015}, the bulge was decomposed into inner and main components modelling them with the exponential density profiles.

It should be stressed that one is free to consider various sets of combinations (permutations) of the models within the galaxy. However, for most of them the BIC values were large. Therefore, they have been omitted. Wherever we involve the de Vaucouleurs profile, the BIC was high. As soon as we abandoned the de Vaucouleurs profile and used the exponential density profile, the BIC started decreasing, see Table ~\ref{tab:third} for details.

According \cite{2014MNRAS.443.2204P} the median mass of the Andromeda galaxy, including the dark matter mass, is around $(1.5\pm0.5)\times10^{12}M_\odot$. \cite{2006ApJ...653..255C} using NFW profile in their analyses, indicate the size of the Andromeda galaxy to be $\sim 70$ kpc in diameter, so $\sim35$ kpc in radius. These data can be used as additional references for our findings. Thus, the results obtained in Table~\ref{tab:third} for the Brownstein and NFW profiles are consistent with the ones in the literature. However, from the statistical point of view the exponential density profile turned out to be the best.

Using updated data points for the RC given by \cite{Sofue2015}, similar analyses have been carried out also for the Milky Way galaxy. Here, the two bulges model with the exponential sphere profile was better than a single bulge model  with the de Vaucouleurs profile. For the halo region,  the exponential sphere profile again provides  statistically the best fit.

In Table~\ref{tab:comparison}, we show the best fit parameters for the two galaxies. It turned out that the Milky Way is roughly twice less massive than the Andromeda galaxy. These results have been compared with those given in Table 2 of Ref. \citet{Sofue2015} and in Table 6 of Ref. \citet{2017PASJ...69R...1S}. It appeared that the order of the parameters are in agreement with \citet{Sofue2015} and \citet{2017PASJ...69R...1S}. The slight differences are due to different approaches employed to fit the data and models chosen for the galaxy halos. As one can see from the Tables, the halo profile affects the inner parts of the galaxies. Therefore, the scale radius, characteristic densities and, as a consequences, the mass vary from the ones reported in the literature.

For both galaxies, we investigated the role of the sound speed as it is closely linked to perturbation theory and structure formation. In particular, it determines the
length above which gravitational instability overcomes the
radiation pressure, where  perturbations can significantly grow. The two galaxies exhibit similar behaviors with different shapes of the rotation curves; this is probably due to the fact that the two galaxies, albeit similar, are morphologically different.

Though one can determine the distribution of dark matter mass in galaxies with the help of rotation curves by assuming various theoretical models, it is not yet clear why dark matter is distributed differently in every galaxy and what is its role in the evolution of the universe. The realization that the nature of the main matter content of the Universe is not well-understood, is a solid motivation to study its properties in future works.

\section*{Acknowledgements}
OL expresses is grateful to the Department of Physics of the Al-Farabi University for hospitality during the period in which this manuscript has been written. This research has been partially funded by the Science Committee of the Ministry of Science and Higher Education of the Republic of Kazakhstan (Grant No. AP08052311).
The work of HQ was partially supported by  PAPIIT-DGAPA-UNAM, Grant No. 114520, and Conacyt-Mexico, Grant No. A1-S-31269.

\bibliographystyle{mnras}

\end{document}